\documentclass{aa}
\usepackage{graphicx}
\usepackage[varg]{txfonts}
\usepackage{natbib,twoopt}
\usepackage[breaklinks=true]{hyperref} 
\bibpunct{(}{)}{;}{a}{}{,} 
\makeatletter
\newcommandtwoopt{\citeads}[3][][]{\href{http://adsabs.harvard.edu/abs/#3}%
{\def\hyper@linkstart##1##2{}%
\let\hyper@linkend\@empty\citealp[#1][#2]{#3}}}
\newcommandtwoopt{\citepads}[3][][]{\href{http://adsabs.harvard.edu/abs/#3}%
{\def\hyper@linkstart##1##2{}%
\let\hyper@linkend\@empty\citep[#1][#2]{#3}}}
\newcommandtwoopt{\citetads}[3][][]{\href{http://adsabs.harvard.edu/abs/#3}%
{\def\hyper@linkstart##1##2{}%
\let\hyper@linkend\@empty\citet[#1][#2]{#3}}}
\newcommandtwoopt{\citeyearads}[3][][]%
{\href{http://adsabs.harvard.edu/abs/#3}
{\def\hyper@linkstart##1##2{}%
\let\hyper@linkend\@empty\citeyear[#1][#2]{#3}}}
\makeatother

\usepackage{lscape}
\usepackage{xcolor}

\begin{document} 
   \title{Convective blueshift strengths of 810 F to M solar-type stars\thanks{Table \ref{tab:cheatsheet} and the full Table \ref{tab:star_data} are available in electronic form at the CDS via anonymous ftp to \url{cdsarc.u-strasbg.fr} (130.79.128.5) or via \url{http://cdsweb.u-strasbg.fr/cgi-bin/qcat?J/A+A/}}}


   \author{F. Liebing\inst{\ref{inst1}} \and S. V. Jeffers\inst{\ref{inst2}} \and A. Reiners\inst{\ref{inst1}} \and M. Zechmeister\inst{\ref{inst1}}}

   \institute{Institut f\"ur Astrophysik (IAG), Universit\"at G\"ottingen,
              Friedrich-Hund-Platz 1, 37077 G\"ottingen, Germany\\
              \email{florian.liebing@uni-goettingen.de}\label{inst1}
              \and
              Max Planck Institute for Solar System Research,
              Justus-von-Liebig-weg 3, 37077 G\"ottingen, Germany\label{inst2}
	     }

   \date{Received xxx /	Accepted xxx}
 
  \abstract
   {The detection of Earth-mass exoplanets in the habitable zone around solar-mass stars using the radial velocity technique requires extremely high precision, on the order of 10\,cm\,s$^{-1}$. This puts the required noise floor below the intrinsic variability of even relatively inactive stars, such as the Sun. One such variable is convective blueshift varying temporally, spatially, and between spectral lines.}
   {We develop a new approach for measuring convective blueshift and determine the strength of convective blueshift for 810 stars observed by the HARPS spectrograph, spanning spectral types late-F, G, K, and early-M. We derive a model for infering blueshift velocity for lines of any depth in later-type stars of any effective temperature.}
   {Using a custom list of spectral lines, covering a wide range of absorption depths, we create a model for the line-core shift as a function of line depth, commonly known as the third signature of granulation. For this we utilize an extremely-high-resolution solar spectrum (R$\sim$1\,000\,000) to empirically account for the nonlinear nature of the third signature. The solar third signature is then scaled to all 810 stars. Through this we obtain a measure of the convective blueshift relative to the Sun as a function of stellar effective temperature.}
   {We confirm the general correlation of increasing convective blueshift with effective temperature and establish a tight, cubic relation between the two that strongly increases for stars above $\sim$5800\,K. For stars between $\sim$4100\,K and $\sim$4700\,K we show, for the first time, a plateau in convective shift and a possible onset of a plateau for stars above 6000\,K. Stars below $\sim$4000\,K show neither blueshift nor redshift. We provide a table that lists expected blueshift velocities for each spectral subtype in the data set to quickly access the intrinsic noise floor through convective blueshift for the radial velocity technique.}
  {}


   \maketitle
%

\section{Introduction}
The vast majority of exoplanets that have been discovered to date using the radial velocity (RV) method have been detected using instruments with a precision of 1\,m\,s$^{-1}$. They have almost exclusively been detected orbiting solar-type stars, defined as stars possessing a convective envelope (i.e., spectral types of late F and below). Expanding the search to lower-mass exoplanets around solar-type stars remains a highly challenging but rapidly progressing field, both technologically and analytically. Examples for instruments that have been successfully employed to this end include CARMENES \footnote{Calar Alto high-Resolution search for M dwarfs with Exoearths with Near-infrared and optical \'Echelle Spectrographs} \citepads{2016SPIE.9908E..12Q} and HARPS \footnote{High-Accuracy Radial velocity Planetary Searcher} \citepads{2003Msngr.114...20M} among many others. The next generation of instruments has been designed to reach the 10\,cm\,s$^{-1}$ level of precision; these include ESPRESSO \footnote{Echelle SPectrograph for Rocky Exoplanets and Stable Spectroscopic Observations} \citepads{2010SPIE.7735E..0FP} and EXPRES \footnote{EXtreme PREcision Spectrometer} \citepads{2016SPIE.9908E..6TJ}, which have recently been commissioned, and others such as ELT-HIRES \footnote{Extremely Large Telescope-HIgh spectral REsolution Spectrograph} \citepads{2016SPIE.9908E..23M}, which are expected further in the future. Instrumentation of this level is required in order to detect Earth-like planets around solar-type stars, which, using the RV method, requires an instrumental precision of 10\,cm\,s$^{-1}$. However, extreme-precision instrumentation only solves part of the challenge.\par
Stellar surfaces, particularly those with an underlying convection zone, are not featureless and exhibit strong spatial and temporal variations. Well-known examples for the effects of magnetic activity include starspots and plages that corotate with the star and periodically modulate the RV on the order of several m\,s$^{-1}$ \citepads{2015ApJ...812...42B}. This is due to asymmetries in the shape of spectral lines, introduced by magnetically active regions deviating from the quiet photospheres temperature, impacting the very precise measurement of the line center required for the RV technique. On a lower level, the statistical nature of stellar surface convection becomes relevant for Earth-twin detections. Granulation, including supergranulation, is one of the defining traits of solar-type stars; this is due to their outer convective zone and varies on timescales of minutes to days. It introduces further RV variations on the order of 10\,cm\,s$^{-1}$ up to 1\,m\,s$^{-1}$, respectively, eclipsing any potential ultra-short-period, sub-Earth-mass planets and hiding the signal of an Earth-twin under stellar jitter. Even higher jitter levels are introduced through the suppression of convective motion by magnetic activity (e.g., \citeads{2014MNRAS.438.2717J}; \citeads{2018A&A...610A..52B}; \citeads{2010A&A...512A..39M}).\par
Another related jitter source is stellar acoustic oscillations, stochastically excited by turbulent convective motion. For stars such as the Sun, the amplitude of the mode of highest power is $\sim$20\,cm\,s$^{-1}$ on a timescale of 5 minutes (\citeads{1995A&A...293...87K}; \citeads{2008ApJ...682.1370K}), even when neglecting all other modes. These oscillations can be mitigated by carefully choosing integration or coaddition times to match stellar oscillation frequencies (\citeads{2019AJ....157..163C}; \citeads{2011A&A...525A.140D}). With all these jitter sources operating on a level comparable to the expected planetary signal but otherwise unrelated timescales, understanding the nature of each individual source and the effect it has on the overall velocity measurement is paramount in order to model and remove their influence (e.g., \citeads{2017A&A...607A...6M}; \citeads{2020ApJ...888..117M}).\par
In this paper we aim to measure one such contribution, the base level of convective blueshift (CBS), for a sample of >800 stars spanning spectral types from late-F to early-M. We expect a strong dependence of CBS on spectral type for two reasons: The depth of the convection zone increases and the energy flux decreases with decreasing mass. These fundamental changes in convective properties should be reflected in CBS. We begin with a description of the general principle behind convective line shift in Sect. \ref{sec:conv_shift_theory}. Section \ref{sec:data} details the data set that forms the foundation of this work, with Sect. \ref{sec:data_proc} explaining the preparatory steps required on the data. The origin, processing, and results of the solar reference data are discussed in Sect. \ref{sec:reference_data} and the final analysis of the sample stars with their results in Sect. \ref{sec:third_sig_fit}. The results are discussed in Sect. \ref{sec:discussion} and summarized in Sect. \ref{sec:conclusion}.

\section{Convective line shift}
\label{sec:conv_shift_theory}
\subsection{Flux asymmetries from granulation}
\label{subsec:flux_asym}
The surfaces of cool stars, meaning those with an outer convective envelope, characteristically show a granulation pattern. It is composed primarily of granules, wide regions wherein hot material rises to the top of the convection zone, cools off, moves to the sides, and sinks back down between the granules in narrow intergranular lanes \citepads{2012AJ....143...92G}. These lanes form the secondary part, separating the granules, and appear darker due to the lower temperature of the material therein just as the granules appear brighter from the hotter material. In the case of a spatially unresolved star where only the disk-integrated stellar flux can be measured, the granules contribute more to the overall flux compared to the intergranular lanes due to their higher brightness and much larger covered area, leading to an imbalance in the flux representation. The magnitude of this imbalance changes strongly with stellar effective temperature, metallicity, and age, all of which influence the global convective pattern in terms of relative area covered by granules, absolute granule size and brightness, vertical velocity, and intergranular lane brightness and coverage, as well as the contrast between the two (empirical: \citeads{1982ApJ...255..200G}, numerical: \citeads{2013A&A...558A..48B}; \citeads{2014arXiv1405.7628M}).\par
\citetads{2014arXiv1405.7628M} show that the dominant granule size increases with temperature and metallicity and significantly decreases with surface gravity. Larger granules in turn are shown to be brighter and hotter and therefore have a higher contrast. The change in peak brightness and temperature with granule size is stronger in hotter or low metallicity stars, but is only minorly affected by surface gravity. Vertical velocity also increases with temperature and metallicity, as well as lower surface gravity, and shows a peaked distribution centered on the mean granule size, with the peak more pronounced in hotter stars.\par
The flux imbalance is unstable over time due to the randomly evolving distribution of granule size and location. The contrast between granules and lanes is constantly changing, and temporary patterns are rotating into and out of view. This limits the RV precision achievable for all measurements relying only on spectral data without accounting for these processes.

\subsection{Impact on line shape}
In convective granulation, rising material contributes more to the shape and location of a given spectral line than the sinking one. As it moves toward the observer, convection leads to a net blueshift. The magnitude of the shift further depends on the depth of formation of the spectral line, since convective velocity changes with depth, which roughly anticorrelates with the lines absorption depth. Therefore, with convection slowing down toward the stellar surface and deeper lines forming at higher layers, deeper lines tend to be less affected by convective line shift. The wings of deeper lines are then shifted independently from the core, following their contribution function, as if they were shallower lines of their own \citepads{2010ApJ...721..670G}. Shallower lines are affected more strongly by convective shifts due to their deeper formation. This leads to an overall asymmetry in the line profile. The degree of this asymmetry then depends on the actual gradient of the convective velocity field with physical depth and the total absorption depth of the line. Lastly, for spatially resolved spectroscopy, meaning only on the Sun for now, limb angle factors in, as observation closer to the disk edges pass through the atmosphere at a shallower angle, observing longer paths while still high in the atmosphere. This increases the optical depth corresponding to a given physical depth and decreases the range of physical depths observable. The line contribution functions, depending on optical depth, shift upward in the atmosphere, changing the line profiles observed depending on limb angle \citepads{2013A&A...558A..49B}. The RV method therefore suffers not only from the constant changes in the convective pattern and therefore the base line shift, it is further complicated by the dependence of overall velocity shift on the choice of spectral lines and method of measuring either each line's individual velocity or a combined value. Conversely, what is a problem for exoplanet hunters can be a major source of information for stellar astrophysicists. 

\subsection{Blueshift measuring techniques}
Both the line asymmetry and differential central shift can be utilized to trace the convective velocity field inside the stellar photosphere. A single, high-resolution spectral line bisector traces convective velocity through the photosphere by its optical-depth dependent contribution function. The line core behaves similarly and, if measured for a large number of lines with differing central depths, can be used to trace velocity through the photosphere as well, with much lower requirements on data quality \citepads{2010ApJ...710.1003G}. Using both approaches simultaneously allows the so-called flux deficit due to the opposing contributions from rising and sinking material to be studied.

\subsubsection{High-resolution techniques: Line bisectors}
\begin{figure}
\resizebox{\hsize}{!}{\includegraphics{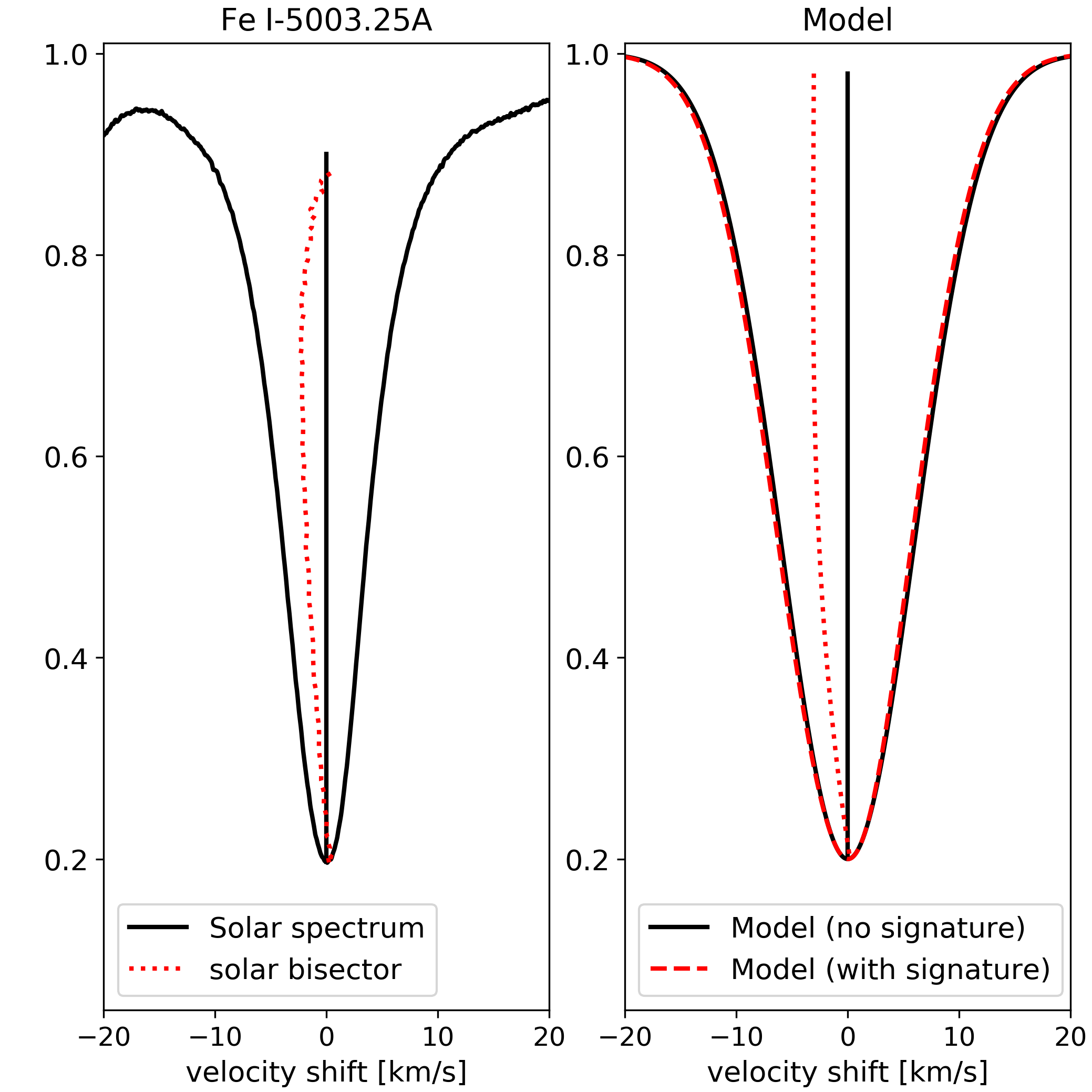}}
\caption{Comparison of a solar and a model spectral line bisector. Left: Solar \ion{Fe}{i} line at 5003.25\,\AA~(solid black) with its center marked by a vertical line (solid black) from the IAG solar flux atlas (used as a template in this work) and the corresponding line bisector (dashed red). Right: Gaussian model line profile without (solid black) and with a solar-like third signature (dashed red, following Eq. \ref{eq:solar_third_sig}). Their respective bisectors are shown in the same colors. The model with the signature is shifted to match the line cores. In both panels, to emphasize the slight blueshift, the bisector scale is amplified by a factor of ten.}
\label{img:Sim_shift}
\end{figure}
One commonly used method to determine the CBS is the line bisector. It is a curve connecting the midpoints of a spectral line along points of equal absorption depth in the red and blue wing. For cool stars the bisector commonly shows a distorted "C" shape, as shown in the right panel of Fig. \ref{img:Sim_shift} for a solar \ion{Fe}{i} line from the \citetads{2016A&A...587A..65R} solar atlas. The foot of the "C," the core of the line, is formed high up in the atmosphere in the overshoot region where convective velocity has decreased significantly. Further up in line flux the contribution function shifts its maximum to higher optical depths with higher convective velocities, forming the bulging part of the "C" shape. At the top of the wings the intergranular contribution becomes significant and adds a redshifted component, counteracting the granular component, which is decreasing in flux. This turns the bisector back toward lower blueshift, completing the "C" shape, as detailed for example in \citetads{1981A&A....96..345D} or \citetads{2019ApJ...879...55C}. This "curve-back" is also known as the flux deficit and symbolizes the difference between a pure granular or third-signature model (see Sect. \ref{subsec:CB_lowres}) to the actual bisector. In effect, the flux deficit extracts and quantifies the intergranular contribution \citepads{2010ApJ...710.1003G}. Figure \ref{img:Sim_shift}, in the left panel, shows as an example one line from the extremely-high-resolution, R$\sim$1\,000\,000 solar spectrum used in this work. A Gaussian model profile on the right demonstrates the third-signature model we created (Eq. \ref{eq:solar_third_sig}). The contribution of the intergranular lane is not within the scope of this work.\par
Properties of these bisectors have been used by various groups (e.g., \citeads{2001A&A...379..279Q}; \citeads{2019ApJ...879...55C}) to characterize the strength of convection, the presence of starspots, limit the RV jitter and, by selecting the lowest region of the bisector, the convective line shift \citepads{1986A&A...158...83D}. It can also be used to delineate the so-called granulation boundary that separates stars with an outer convective envelope from radiative ones. This happens around early-F types on the main sequence with the bisector first loosing the C-shape, straightening out, and then reversing shape \citepads{1989ApJ...341..421G}; a process that has been closely examined in \citetads{2010ApJ...721..670G} and could be explained with a shift in location along the wing of the flux deficit.\par
Using the bisector to determine convective shifts has the advantage that only a handful of lines or even single ones are sufficient to provide an accurate representation of the convective profile. The disadvantages are that it is particularly sensitive to blended lines and line distortions. It also requires spectral lines to be of very high resolution (R$\gtrsim$300\,000) and high signal-to-noise (S/N$\gtrsim$300) \citepads{1987A&A...172..200D}; otherwise, line asymmetries might be smeared out from the wider instrumental profile or lost in noise and photon binning.

\subsubsection{Low-resolution techniques: Third-signature scaling}
\label{subsec:CB_lowres}
In this work, we expanded on a technique to measure the CBS of stars with spectra of a much lower resolution of R$\sim$100\,000 and comparatively low S/N, typically at $\sim$500 for our data, though S/N as low as 100 are sufficient (Sect. \ref{subsec:S/N_effect}). The technique is significantly less demanding compared to single-line bisector analysis and applicable to stars spanning spectral types late-F to early-M. This was achieved with an extremely-high-resolution solar Fourier-transform-spectrograph (FTS) spectrum that was used to create a high-quality template for the magnitude of CBS. This CBS template was scaled to match measurements from lower-resolution, lower-S/N stellar spectra to obtain a relative CBS strength. The backbone of this approach is the switch from individual line bisectors to many lines' core shifts, which allows for the accuracy of ultra-high-quality data in the form of an empirical template combined with the greater availability of lower-quality data. This technique hereby takes advantage of two key properties of CBS.\par
The first advantage comes from utilizing the central shifts for lines of different depths, ideally of the same element and ionization state to avoid additional degrees of freedom, as a tracer for the CBS \citepads{1999PASP..111.1132H}. This enables the usage of a much larger sample of lines, especially for fainter stars with generally insufficient levels of S/N for the bisector technique. The relation between the central line shift and line depth is commonly referred to as the "third signature of granulation", with the first and second being the line broadening due to macroturbulence and the line asymmetry, respectively \citepads{2009ApJ...697.1032G}. This is of further use if the goal is extreme-precision stellar RV determination using line-by-line measurements. Here, the selection of lines becomes important since different lines, through their different depths, have different intrinsic shifts that also change between stars \citepads{2020A&A...633A..76C}.\par
The second advantage of the third-signature technique, when applied to multiple stars, is that it is capable to supply additional information. While the absolute values of CBS of the line cores will change from star to star for similar absorption depths, the basic shape of the third signature is universal (\citeads{2009ApJ...697.1032G}; \citeads{2018ApJ...852...42G}), except for stars above the granulation boundary (around spectral type F0 on the main sequence, just above 7000\,K; cooler post main sequence)\citepads{2010ApJ...721..670G} where the flux deficit, or intergranular lane contribution, deforms the middle part of the third signature instead of the top toward redder shifts. The third signature can be matched between different stars by scaling and shifting the velocities and hence allows a relative convection strength to be determined from the scaling and a relative RV from the shift. To create a template on which to base the signature matching and to provide reference values for blueshift and RV, a single high-resolution, high-S/N spectrum, in our case from the Sun, is sufficient. The derived template for the third signature of granulation is then shifted and scaled to match the data from the other stars. Measuring CBS this way increases the robustness for a given data set due to the much lower requirements on resolving power (Sect. \ref{subsec:R_effect}), the ability to more easily use coaddition to increase the S/N, the generally lower requirement on S/N (Sect. \ref{subsec:S/N_effect}), and the reduced possibility of third-signature fitting errors due to the exact shape being empirically prescribed instead of assumed or simplified. In turn, this requires much greater care in the definition of the template signature because all uncertainties will be amplified through all other stars' analysis, necessitating the use of FTS solar data rather than HARPS standard stars.

\section{Data}

\subsection{HARPS stellar data}
\label{sec:data}
This work is based on data obtained from HARPS, a fiber-fed, cross-dispersed echelle spectrograph installed at the 3.6\,m telescope at La Silla, Chile. The instrument itself is housed in a temperature stabilized, evacuated chamber, and covers a wavelength range from 380\,nm to 690\,nm at a resolving power of R=115\,000 over 72 echelle orders. It is capable of reaching a precision of 1\,m\,s$^{-1}$ \citepads{2003Msngr.114...20M}. The data were composed by \citetads{2020A&A...636A..74T}, who collected and sorted through all publicly available spectra from the HARPS spectrograph.\par
\begin{figure}
\resizebox{\hsize}{!}{\includegraphics{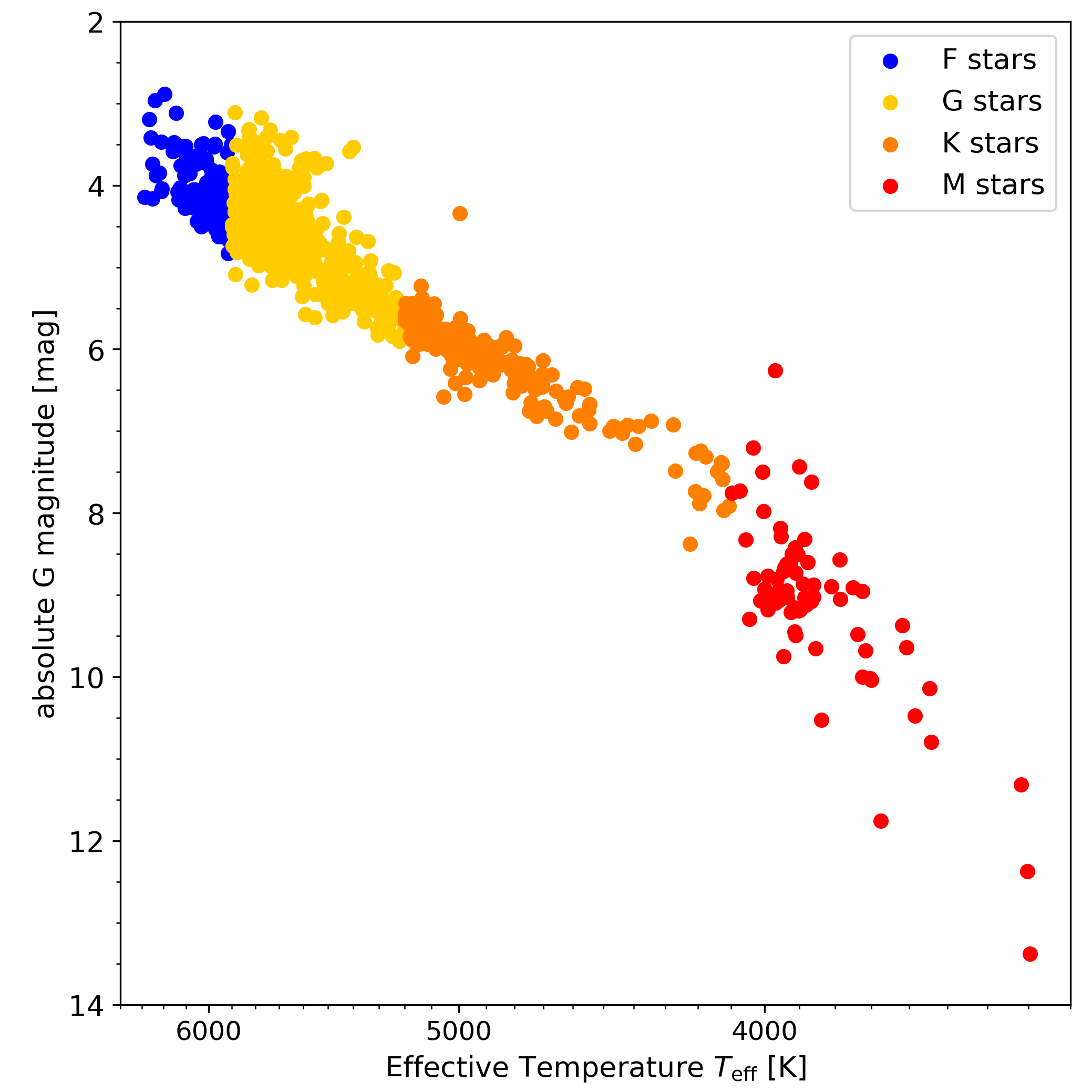}}
\caption{Hertzsprung-Russel diagram of the 810 stars in the HARPS sample, based on GAIA DR2 data.}
\label{img:data_HR}
\end{figure}

The full sample consists of 3094 stars. Out of that number, 439 were excluded because they had no matching counterpart in GAIA DR2 \citepads{2018yCat.1345....0G} and 458 were identified as subgiants. The subgiants were excluded to be investigated in more detail in a later paper. The method used in this work (Sect. \ref{sec:data_proc}) at HARPS resolving power further restricts the usage to stars with a projected rotational velocity of $v \sin i <$ 8\,km/s (Sects. \ref{subsec:R_effect}, \ref{subsec:vsini_effect}), which excluded all stars above 6250\,K. We used $v \sin i$ values from SIMBAD and \citetads{2005yCat.3244....0G} as well as HARPS DRS FWHM values listed in \citetads{2020A&A...636A..74T} as proxy to exclude stars that did not meet this criterion. Filtering also stars of unknown $v \sin i$ left 810 stars in our final sample for analysis, that span the range from early-M to late-F types. The location of the 810 sample stars on the Hertzsprung-Russel diagram, based on GAIA DR2 data, is shown in Fig. \ref{img:data_HR}.\par
\citetads{2020A&A...636A..74T} coadded and analyzed the spectra for each star using the "SpEctrum Radial Velocity AnaLyser" ({\tt serval}, \citeads{2018A&A...609A..12Z}), corrected for nightly offsets, and provided an overhauled RV data set. We used the high-S/N, coadded, template spectra, created by the {\tt serval} pipeline for its template matching approach to RV determination, as well as the final radial velocities. During coaddition, {\tt serval} further corrected any long-term trends in RV, such as binary motion.\par
\begin{figure*}
\resizebox{\hsize}{!}{\includegraphics{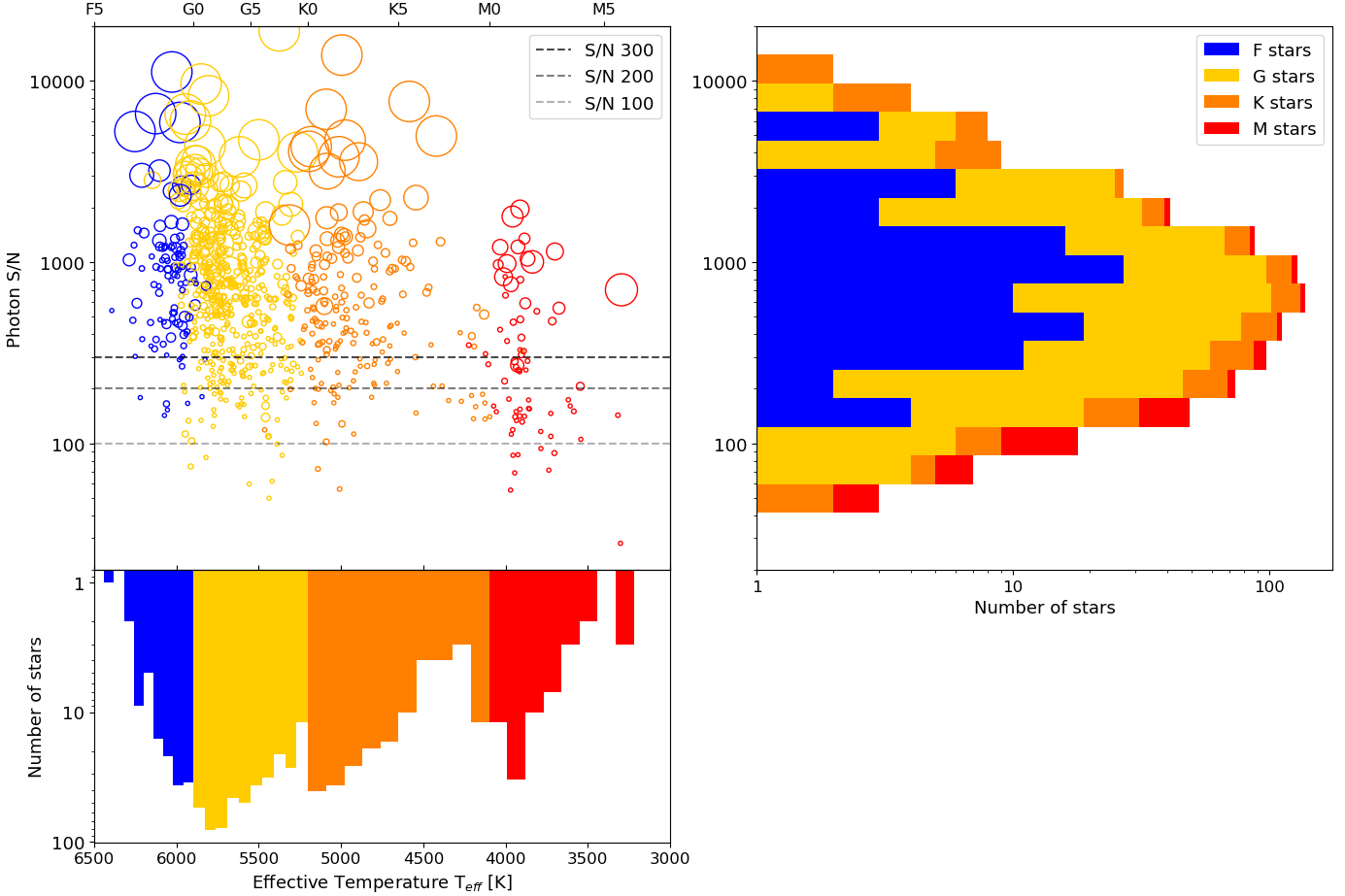}}
\caption{Overview of all sample stars' signal-to-noise ratio (S/N), temperature, and spectral type. Top left: Distribution of S/N over spectral type within the sample. The marker size corresponds to the number of observations for that star. The M stars show significantly smaller S/N with median below 300 due to both smaller intrinsic brightness and lower average number of coadded spectra. The dashed, horizontal lines mark S/N values of 100, 200, and 300, motivated by the results from Sect. \ref{subsec:S/N_effect}. Top right: Peak S/N for the coadded spectra in the sample. Bottom left: Distribution of spectral types within the sample.}
\label{img:data_S/N_hist}
\end{figure*}

The {\tt serval} coadded spectra are given in vacuum wavelength and were converted by us to air wavelengths following the method from \citetads{1996ApOpt..35.1566C} as given in \citetads{2013A&A...553A...6H}. This is necessary because the line-sets' wavelengths are given in air (Sect. \ref{subsec:line_select}). The S/N ratios obtained by {\tt serval} from the coaddition of the spectra are shown in Fig. \ref{img:data_S/N_hist} in the top panels for each star individually (left) and binned by S/N (right, stacks colored to reflect spectral type). On average, 66 spectra were used per coaddition for an approximate S/N just below 570, with some stars significantly higher. The lower panel of Fig. \ref{img:data_S/N_hist} shows the sample size, binned roughly by spectral subtype approximated by and plotted in temperature. The actual spectral subtype used to determine temperature-bin edges was approximated with GAIA (BP-RP) colors, converted to SDSS (g-i)\footnote{\url{https://gea.esac.esa.int/archive/documentation/GDR2/Data_processing/chap_cu5pho/sec_cu5pho_calibr/ssec_cu5pho_PhotTransf.html}}, and then interpolated following \citetads{2007AJ....134.2398C}. The sample misses some late-K type stars and very-high-S/N M-types. The latter is unsurprising due to the low intrinsic brightness of those stars and the former a known deficiency in classical spectral typing that leads to the subtypes of K8 and K9 to be virtually nonexistent. Despite this, the sample is continuous in temperature. A complete list of the stars from the final sample with their parameters can be found in Table \ref{tab:star_data} (full version available at CDS).

\subsection{IAG solar data}
\label{subsec:IAG_data}
To create the high-precision third-signature model required for the HARPS scale factor determination, we used the IAG solar FTS atlas\footnote{\url{http://www.astro.physik.uni-goettingen.de/research/flux_atlas/}} \citepads{2016A&A...587A..65R}. With a spectral range of 405-2300\,nm it covers the entire visible range and, most importantly, the entire range of HARPS data used in this work enabling the use of matching line lists. The solar spectrum was recorded with an FTS over 1190 scans in the VIS range at a resolving power of over R$\sim$1\,000\,000, exceeding HARPS data by an order of magnitude. This extreme data quality ensures we are only limited by the algorithm itself, variabilities intrinsic to the stars, and remain independent of any specific model assumptions.

\section{HARPS data processing}
To measure the CBS, we normalized the continuum of the spectra and created a preliminary list of spectral lines we used for the third-signature fit. The absolute stellar RV was obtained from the \citetads{2020A&A...636A..74T} RVBANK as a reference. In an iterative process the preliminary line list and RV were further refined.
\label{sec:data_proc}
\subsection{Boundary fit}
The first step in measuring CBS is to normalize the continuum level, correcting for three effects: The instrumental blaze function, the gradient from the blackbody spectrum, and molecular bands for cool stars. We assumed that both the blackbody contribution and molecular bands could be neglected within single diffraction orders compared to the blaze function. To remove the latter and normalize the continuum, an upper boundary fit was employed, generally following \citetads{2009MNRAS.396..680C}; however, we used a squared sine-cardinal function to mimic the blaze function. This differs from the usual approach of binning the spectrum within an order by wavelength, estimating a maximum as the local continuum, and interpolating the binned continuum levels. This approach, while functional, is highly susceptible to nonlocal spectral features, for example absorption bands, as well as emission lines, relies on a good choice of interpolation, and neglects previous knowledge about the continuum shape. The cost function, again generally following \citetads{2009MNRAS.396..680C}, was minimized with a Nelder-Mead simplex algorithm implementation from the python package {\tt scipy.optimize.minimize}. To account for strong absorption (e.g., Na-D) and strong emission lines (e.g., Ca H+K), the algorithm presented by \citetads{2009MNRAS.396..680C} was modified to further include an asymmetric kappa-sigma clipping on the residuals to exclude such influences. Lastly, by definition of the boundary-fit algorithm, some valid data points still lie above the fit as the fit is only pushed toward the boundary through asymmetric weighting of the residuals, not onto the actual boundary. Therefore, after normalizing the order first with the fit, those points were assumed to represent the true continuum and the entire order was normalized accordingly. An example for one echelle order is given in the Appendix in Fig. \ref{img:boundfit_example} to illustrate the fit and normalization result.

\subsection{Line selection}
\label{subsec:line_select}
The line list for the third-signature fit must contain lines that are present in all stars from spectral type mid-F to early-M. This ensures that the results remain comparable among themselves and against other choices of lines. The initial lists of atomic and molecular lines were taken from the "Vienna Atomic Line Database\footnote{\url{vald.astro.uu.se}}" (VALD; \citetads{1995A&AS..112..525P}; \citetads{2000BaltA...9..590K}; \citetads{2015PhyS...90e4005R}). The parameters that were used are listed in Table \ref{tab:vald_params}. From the extracted lists we selected only lines with an absorption depth of at least 10\%. Furthermore, only lines with a distance of at least 1\,picometer to their nearest neighbor were selected. This was enforced by an iterative process that always kept the deeper member of the closest line pairs until all remaining lines were sufficiently separated. In this we assumed that line blends, averaged over all lines, did not significantly affect our results. We investigated possible effects from this by comparing average deviations between measurements in wavelength regions containing more or less blended lines. Fits of the CBS in either case do not show significant differences, indicating line blends are not an issue in this instance. This matches findings by \citetads{2009ApJ...697.1032G}, that line core measurements are much more robust against blends compared to bisectors. They further mention the possibility that the cores of very deep lines may no longer probe the photosphere, but instead reach the chromosphere with its wealth of NLTE and magnetic effects. Our analysis does not find qualitative differences in the scaling behavior of lines up to depths of 0.95, extending \citetads{2009ApJ...697.1032G} findings by about 0.2 units in depth.\par
The final reduction of the line list took place after one full analysis run was completed, following the steps outlined in the rest of this section and the main analysis steps given in Sect. \ref{sec:third_sig_fit}. This step was implemented to further remove lines that appeared more sensitive to measurement errors, based on the consistency of their results over all stars. Details on this process are given in Appendix \ref{sec:apdx_linerefine}.
The line lists for different effective temperatures were compared via their results to ensure the choice of temperature did not affect the results significantly (Section \ref{subsec:Teff_effect}). The prescribed temperature shows no significant impact and the main analysis was based on the 3750\,K line set, which, after refinement, comprises 1256 remaining lines, an order of magnitude above the lists used by other groups, for example \citetads{2017A&A...597A..52M}.
\begin{table}
\centering
\caption{Parameters used for the VALD extract stellar query.}
\label{tab:vald_params}
\begin{tabular}{c c c c c}
\hline\hline
$T_{\rm eff}$ & log\,$g$ & $v_{\rm mic}$ & composition & range\\
\hline
3750 K & 5.0 & 1 km\,s$^{-1}$ & solar & 4000-10000 A\\
4500 K & 4.5 & 1 km\,s$^{-1}$ & solar & 4000-10000 A\\
5500 K & 4.5 & 1 km\,s$^{-1}$ & solar & 4000-10000 A\\
6000 K & 4.5 & 1 km\,s$^{-1}$ & solar & 4000-10000 A\\
\hline
\end{tabular}
\end{table}

\subsection{Telluric contamination}
Besides line blends, contamination by telluric absorption lines can impact the accuracy of line core measurements. We used the atmospheric transmission spectrum included in the {\tt ENIRIC} package \citepads{2016A&A...586A.101F}, based on {\tt TAPAS} synthetic spectra \citepads{2014A&A...564A..46B}, to determine bands of potential contamination and assessed their impact on our overall results. Out of the finalized list of lines, only 6\% are in danger of contamination, the majority of them along with the potentially strongest contaminated ones are located toward the red end of the HARPS wavelength range. Similar to line blends, we find no dependence of line scatter on wavelength and excluding the telluric bands does not improve our results. Comparing residuals of the lines in question to lines of similar depth outside the telluric bands again shows no significant differences. From this we determine that telluric contamination can be neglected and continued with the complete reduced line set.

\subsection{Initial radial velocity}
\label{subsec:ccf_rv}
The {\tt serval} pipeline coadds the spectra using differential RVs. In addition to the values from the HARPS RVBANK, we measured the absolute RV of the coadded spectrum using a binary line mask from the CERES pipeline \citepads{2017PASP..129c4002B}\footnote{\url{https://github.com/rabrahm/ceres/tree/master/data/xc_masks}}. This mask is based on an M2 star. We cross-correlated the box function created from the wavelengths and weights from the CERES mask with the spectrum utilizing the {\tt scipy.signal.correlate} package, each with their mean subtracted beforehand, and normalized with the integrated area of the box-function. This cross-correlation function (CCF) is calculated for each order of the coadded spectrum of each star, represented with a cubic spline interpolation, and sampled on a common velocity grid using {\tt scipy.interpolate.splrep/splev}. The CCFs were averaged over the orders and the minimum determined using a fourth order {\tt InterpolatedUnivariateSpline} again from {\tt scipy.interpolate} and its {\tt derivative.roots} function within $\pm$500\,km\,s$^{-1}$. This position was taken as the initial RV.

\subsection{Measuring line positions}
\label{subsec:line_pos_measure}
As an initial RV guess we used the cross-correlation and RVBANK value. The {\tt serval} coadded spectrum of each star was shifted accordingly and the reduced list of lines from Sect. \ref{subsec:line_select} applied as initial line positions. The rest of the paper used the CCF RV correction with the RVBANK values used as validation. Following \citetads{2016A&A...587A..65R}, a parabola was fitted within $\pm$1500\,m\,s$^{-1}$ to the line core and its minimum taken as the center of the line. We refitted the parabola with the new line center until convergence was reached (10$^{-5}$\,nm correction step) in case the initial position was too far from the line core. The final minimum was taken for the line position and absorption depth. The precision in depth, following Sect. \ref{subsec:acc_line_depth}, is on the order of $7 \cdot 10^{-3}$, while Sect. \ref{subsec:acc_line_center} shows a shift accuracy of better than 40\,m\,s$^{-1}$ for single lines. If the fit got stuck in a local maximum, for instance from remaining blends or insufficient RV correction, the last step was repeated with increasingly larger steps away from the maximum added to the last position. Very large steps between iterations were bounded to 1\,pm to prevent fitting errors in shallow, noisy lines that could point the assumed center at far distant wavelengths. Finally, remaining blends that pushed measurements of two lines into the same minimum were taken to be primarily composed of the lesser shifted component with the other removed from the results. Duplicate measurements due to echelle order overlap were averaged in location and depth.

\subsection{Refining radial velocity}
\label{subsec:rv_refinement}
As a last step, the RV was further refined by utilizing the correlation of absorption depth with CBS. Using only the CCF based RV from Sect. \ref{subsec:ccf_rv} or RVBANK value, a first determination of the CBS was done from the line-by-line RV from Sect. \ref{subsec:line_pos_measure}. Since the deepest lines are barely affected by CBS, lines with an absorption depth above 0.9 were then binned in depth and their median RV added to the result from the previous step. Since these lines should show close to zero CBS due to their low depth of formation, they are a good indicator for the overall RV. This cycle of applying RV, measuring CBS, and reevaluating the RV from the deepest lines was repeated until convergence. This additional step was necessary because many lines included in the list are very narrow and therefore require a very good RV correction in order for the initial starting point of the measurements to still lie on their wings and not get shifted to a neighboring line. Using the deep lines to determine the RV is a simple and robust approach.

\section{Modeling the solar third signature}
\label{sec:reference_data}
Before we calculated our third-signature solar template, we first ensured the accuracy of our line center determination. We recomputed the third signature of granulation for the Sun using the list of \ion{Fe}{i} lines from \citetads{1994ApJS...94..221N}, compared the results to \citetads{2016A&A...587A..65R}, before the fit was repeated using our custom line list as explained in Sect. \ref{subsec:line_select}.\par
\begin{figure}
\resizebox{\hsize}{!}{\includegraphics{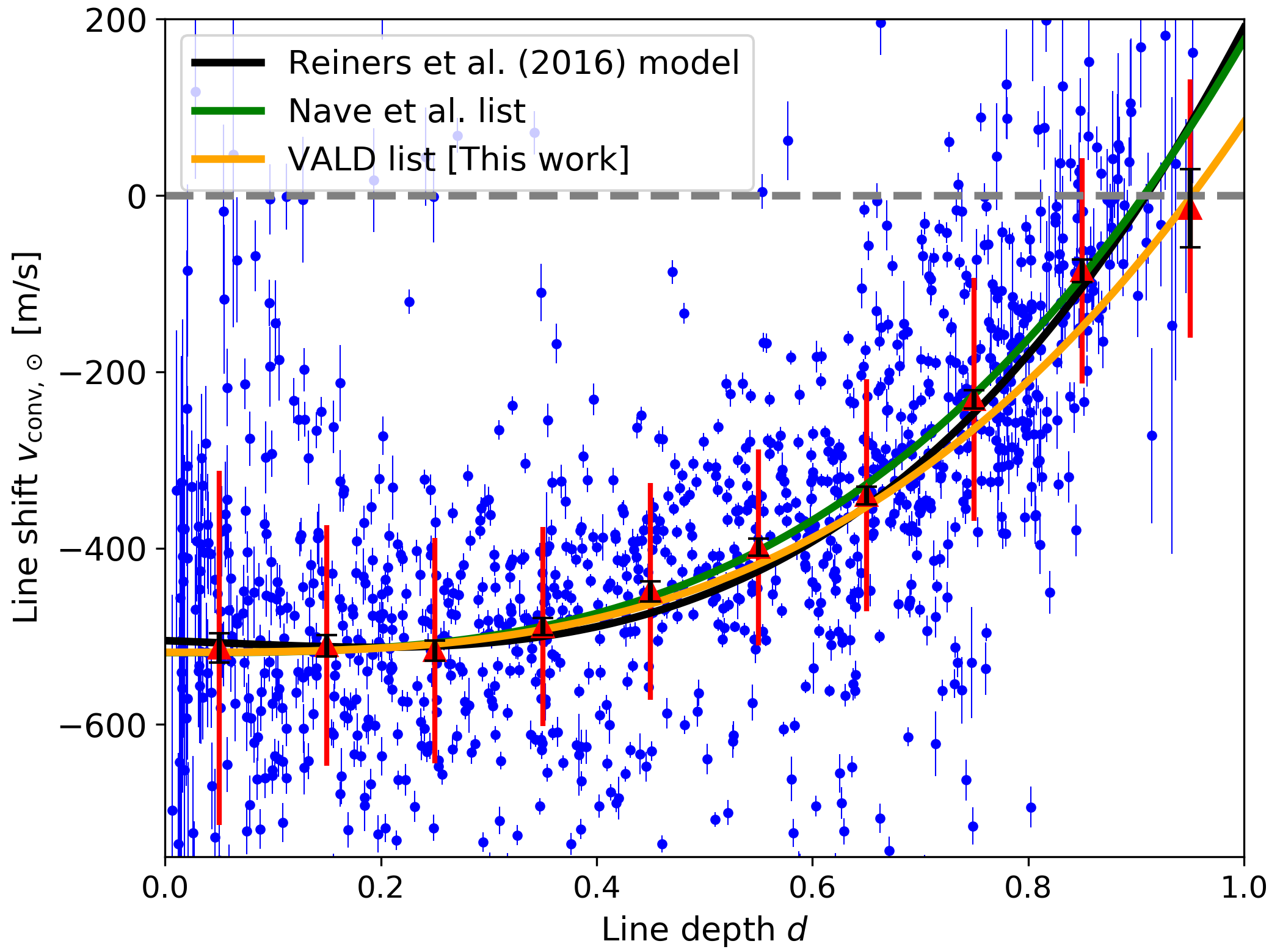}}
\caption{Third signature of granulation extracted from the IAG solar flux atlas. Blue dots with error bars mark the lines measured using the \citetads{1994ApJS...94..221N} list, red triangles with bars are bin medians with MAD, black error bars show the error of the median. The black line is the model for the solar third signature from \citetads{2016A&A...587A..65R} based on the \citetads{1994ApJS...94..221N} line list and the green line is the model from this work (Eq. \ref{eq:solar_third_sig}) based on the same. The orange line is our VALD-lines based template (Eq. \ref{eq:template_third_sig}), shifted to match the intersection of the other two models at depth 0.15.}
\label{img:solar_sig}
\end{figure}

We followed the general procedure from \citetads{2016A&A...587A..65R}: We first measured the line-by-line CBS as the difference between the measured line position and the tabulated rest position. Then the shifts were median-binned by absorption depth and the bin dispersion was characterized with the median absolute deviation (MAD) from {\tt astropy.stats.mad\_std} (Fig. \ref{img:solar_sig}). The error of the bin median was derived from dividing the MAD by $\sqrt{N}$, with $N$ the number of lines in the bin, analogous to the standard error of the mean. This accounts for the significantly higher bin dispersion as opposed to the formal uncertainties on the line shifts. The binned data were fitted with a cubic polynomial, similar again to \citetads{2016A&A...587A..65R}, using the {\tt scipy.optimize.curve\_fit} function in Levenberg-Marquardt mode. This algorithm was used for all other fits as well, unless specifically stated otherwise, and further provided the uncertainties on the fitted parameters.\par
We assumed the following constraints on our fit: The shallowest lines have formed deep enough inside the convection zone for the velocity to be constant. Hence the third signatures' gradient at zero absorption depth was set to zero, eliminating the linear term. Furthermore, we assumed that within the convection zone the velocity only decreases toward the surface. To that end, we ensured a nonnegative gradient of the function. In practice, this led to a quadratic term that approached zero and it was eliminated as well. These assumptions were enforced to remedy a local minimum in the shallowest lines that occurs in \citetads{2016A&A...587A..65R}. This would indicate an outward directed increase in convection velocity, which should only happen at the bottom of the convection zone, far below the line forming region. Our physically motivated constraints avoid this numerical artifact. The resulting polynomial $v_{\rm conv, \odot}\left(d\right)$ for the \citetads{1994ApJS...94..221N} line list is given in Eq. \ref{eq:solar_third_sig}.\par
\begin{align}
v_{\rm conv, \odot}\left(d\right) = 694.325 \cdot d^3 - 518.419,\label{eq:solar_third_sig}\\
v_{\rm conv, \odot}\left(d\right) = 601.110 \cdot d^3 + 173.668.\label{eq:template_third_sig}
\end{align}
The template third signature of granulation used for the remainder of this work was created in two steps: First, we created a base template from full-resolution FTS measurements. Second, the base template was calibrated to the resolving power of the target instrument, in this instance HARPS.\par
The first step is identical to the approach to Eq. \ref{eq:solar_third_sig}, switching from the \citetads{1994ApJS...94..221N} to the VALD line list, while the later follows Sect. \ref{sec:third_sig_fit}. It is necessary to account for the changes in observed line depth introduced through broader instrumental line shapes at lower resolving powers if one expects solar-strength convection to be represented by a solar-relative factor of one or intends to compare results from spectra of different resolving powers. As a relative measure of convection strength using a single instrument, the later step can be skipped. The reason we did not use the iron list from \citetads{1994ApJS...94..221N} for the rest of this work, though it agrees well for the solar Atlas, is because it was compiled specifically for the Sun and did not provide satisfactory results for stars of different spectral types. While our list still mainly consists of \ion{Fe}{i} lines, they are more numerous and diverse, avoiding the over-adaptation to the solar case.\par
Equation \ref{eq:template_third_sig} gives the base template from the first step, created from very-high-resolution FTS data. We degraded the FTS spectrum to a resolving power approximating 110,000 and performed a fit of the base template that results in calibration values of 0.91 in scale and 21.2\,m\,s$^{-1}$ offset. These corrections were implicitly applied for the remainder of this work, whenever the template is mentioned. Additional corrections for different resolving powers can be taken from Fig. \ref{img:solarRTest}.\par
Figure \ref{img:solar_sig} shows the line measurements using the solar atlas data for the \citetads{1994ApJS...94..221N} lines, the binned results, and our fitted polynomials. We confirm our recomputed third-signature template largely matches the results from \citetads{2016A&A...587A..65R} for their line list, except for their local minimum at approximately 0.3 line depth. The signature based on the VALD list deviates for deeper lines and shows a shallower overall shape from the difference in selected lines compared to our fit to the \citetads{1994ApJS...94..221N} lines. The \citetads{2016A&A...587A..65R} signature appears slightly steeper. The R$\approx$110,000 calibrated template is slightly shallower still, as indicated by the smaller than unity calibration factor, clearly demonstrating the necessity of this step for comparable results. The scatter in this work appears slightly smaller on average compared to the results from \citetads{2016A&A...587A..65R} but slightly larger toward the shallower lines.\\
For the template signature used in the remainder of this work, the calibrated, VALD based one, another offset of 726\,m\,s$^{-1}$ was subtracted in all illustrations to adjust the signature such that a fully absorbing line of depth 1.0 corresponds to a CBS of zero. The reasoning is identical to the residual RV correction from Sect. \ref{subsec:rv_refinement} and purely cosmetic to make comparisons between panels easier. The origin of this offset is primarily the solar gravitational redshift of 636\,m\,s$^{-1}$. The remainder is due to the theoretical, fully absorbing line we used as a reference being deeper than the median line used for the 636\,m\,s$^{-1}$ determination.\par
The remaining scatter can be seen for example in \citetads{2016A&A...587A..65R}, \citetads{1999PASP..111.1132H}, and \citetads{1998A&AS..129...41A}, although they used equivalent widths instead of line depths, \citetads{2017A&A...597A..52M}, and \citetads{2017A&A...607A.124M}. A possible correlation with wavelength was explored in, for instance, \citetads{1981A&A....96..345D} or \citetads{1999PASP..111.1132H}, although \citetads{1998A&AS..129...41A} disagree with their conclusion. Our investigation into that topic using the present solar data shows a strong decrease in CBS toward longer wavelengths at similar absorption depths (see Appendix \ref{sec:apdx_lam_effect}). The dispersion visible for lines deeper than $\sim$0.7 is fully explained this way (Fig. \ref{img:solar_sig_interval}), with shallower lines showing additional dispersion.

\section{Third-signature fit for the HARPS sample}
\label{sec:third_sig_fit}
\begin{figure*}
\resizebox{\hsize}{!}{\includegraphics{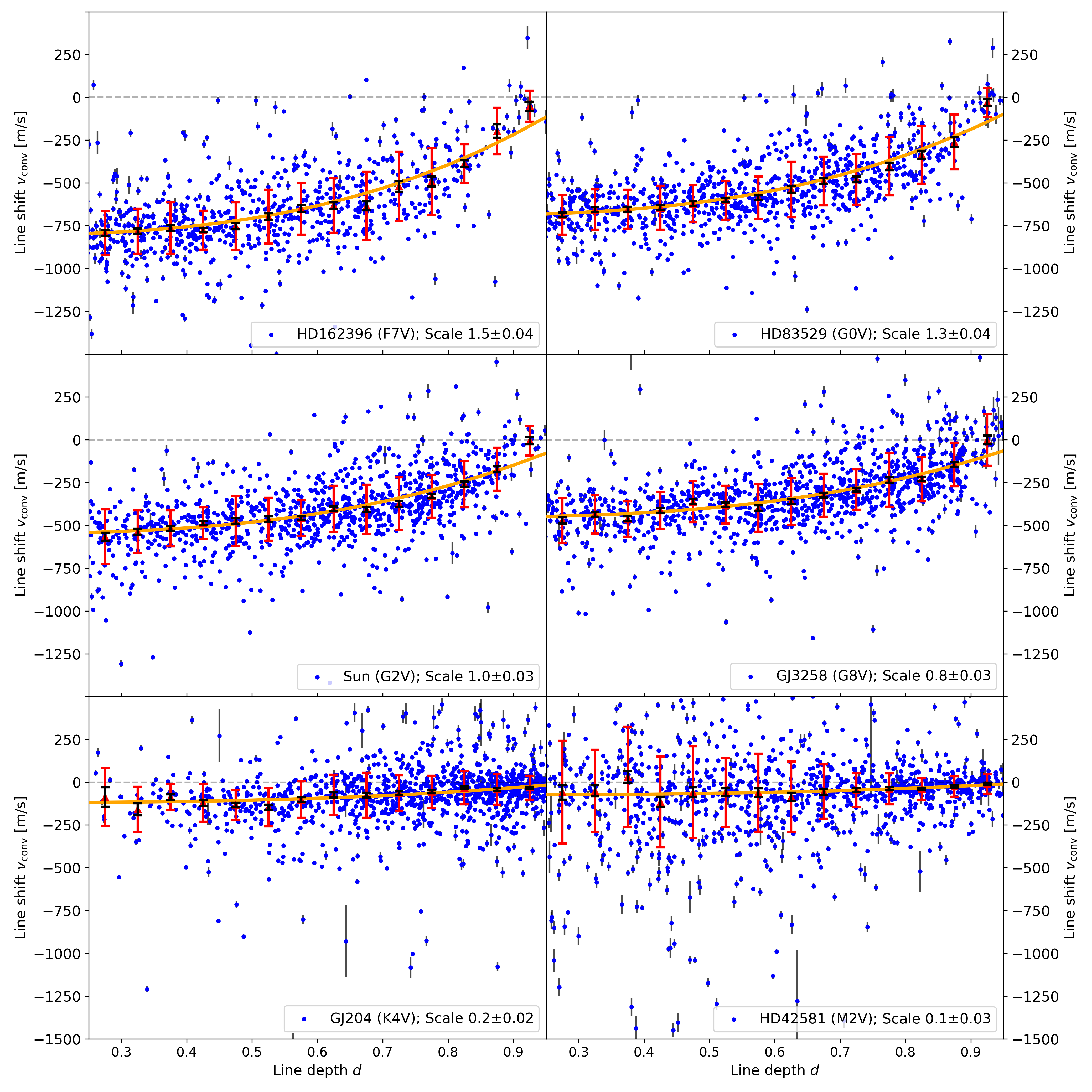}}
\caption{Example third-signature plots derived from HARPS data and solar FTS, convolved to R=110000, for spectral types ranging from F7V (top left) to M2V (bottom left), chosen to sample the scale factor range. Shown are shifts $v_{\rm conv}$ for individual lines (blue circles) and results binned by line depth (red triangles with error bars for MAD, black for the error of the median). The best-fit, scaled, solar third signature is shown (orange curve, fitted through the red markers).}
\label{img:third_sig_examples}
\end{figure*}

We followed the same procedure as the second step of the template creation from Sect. \ref{sec:reference_data} for the VALD line list to fit the solar template to the HARPS measurements. The line list from Sect. \ref{subsec:line_select} was applied to each coadded spectrum for each of the stars in the HARPS sample. Then, the line-by-line measurements of $v_{\rm conv}$ were median-binned by depth $d$ and fitted with the template signature from Eq. \ref{eq:template_third_sig} as $v_{\rm conv}=S\cdot v_{\rm conv, \odot} + v_{\rm 0}$, using a scaling factor $S$ and velocity offset $v_{\rm 0}$. The shallowest and deepest bin were excluded (see Fig. \ref{img:third_sig_examples}). Uncertainties were provided by the fitting algorithm. From the two parameters only $S$ is of further relevance in this work since it encodes the strength of the CBS relative to the Sun. The shift, an RV on the order of a few m\,s$^{-1}$ and comparable in magnitude to the scatter intrinsic to each stars measurements, is an offset inherent in all lines and therefore unrelated to CBS, which operates on a differential, line-by-line basis. It is therefore assumed to be an uncorrected residual from the RV correction and subtracted from each star for all subsequent plots to provide a common velocity zero point.\par

In the next sections we describe the results of applying the solar third signature of granulation template to the HARPS sample of 800 stars. The results are plotted in Fig. \ref{img:third_sig_examples} for one sample star each for spectral types late-F, mid-G, early-K, late-K, and early-M as well as the Sun for comparison. These stars were selected based on their fitted solar relative scale factor, from now on referred to simply as scale factor, in order to evenly span the range available from the sample and to illustrate the gradual change in third signature with spectral type. The full list of scale factors for each star are tabulated in Table \ref{tab:star_data}, available at CDS, and plotted in Fig. \ref{img:scale_Teff}, showing the gradual increase with effective temperature. The solar panel differs from Fig. \ref{img:solar_sig} in that here the VALD filtered list and corresponding template were used, not the \ion{Fe}{i} list from \citetads{1994ApJS...94..221N}, as well as the broadened FTS spectrum.

\subsection{F, G, and K stars}
\label{subsubsec:FGK_results}
\begin{figure*}
\resizebox{\hsize}{!}{\includegraphics{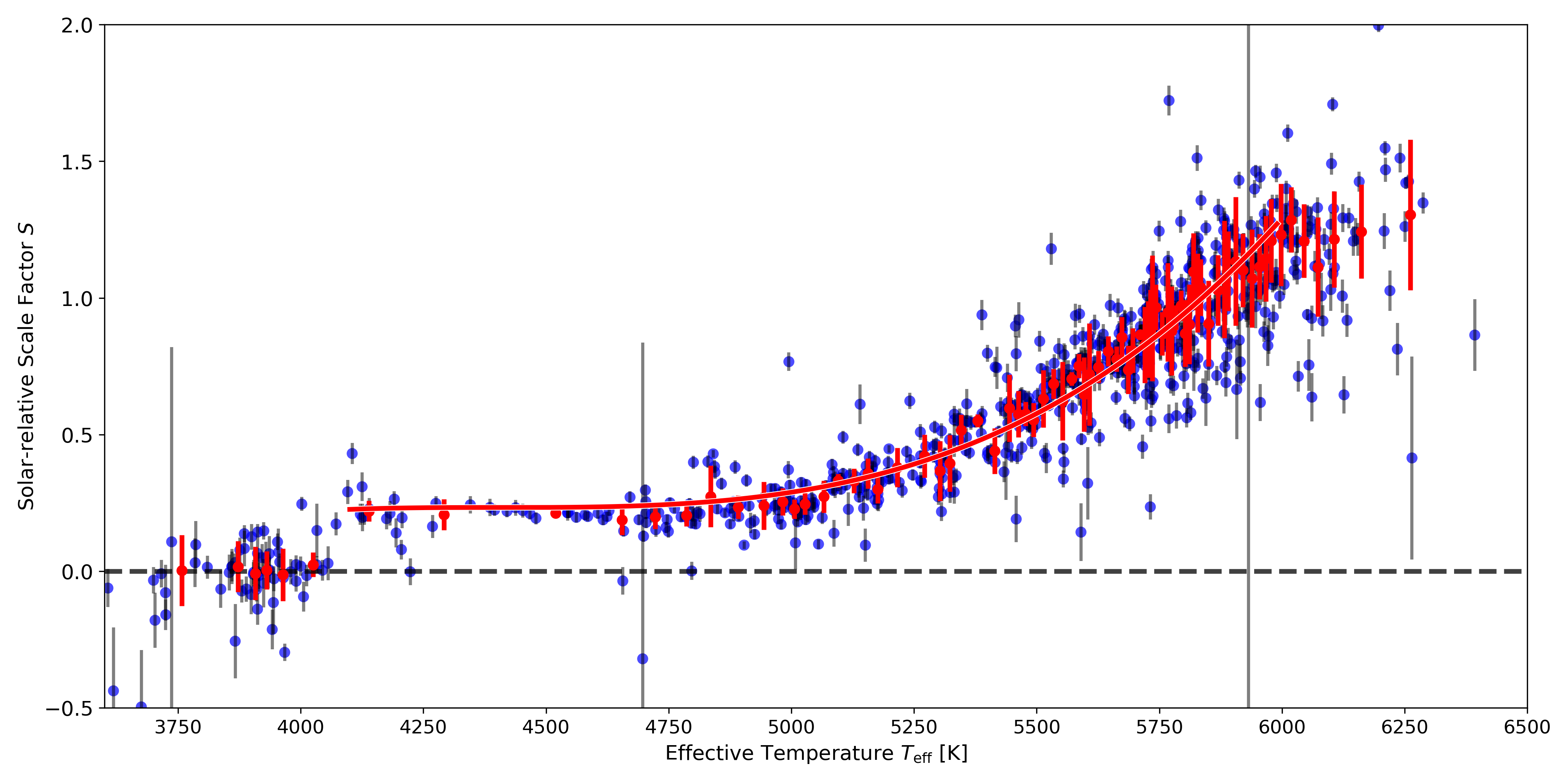}}
\caption{Solar relative scale factor for all main sequence stars in the set (blue circles) plotted against their respective effective temperature. Bins with ten stars each (red circles with error bars) are fitted with a cubic polynomial (red curve, Eq. \ref{eq:scale_model}).}
\label{img:scale_Teff}
\end{figure*}

The results for the F, G, and K dwarfs are shown in Fig. \ref{img:scale_Teff} where there is a strong, increasing dependence between scale factor and effective temperature, which could also be seen in Fig. \ref{img:third_sig_examples}. The main sequence stars increase steeply in scale factor for hotter stars, flatten toward early K-types, and plateau for later K-types. To quantify this relation, the scale factors for the main sequence stars were median-binned by temperature, with the error of the bin median obtained from the MAD as in Sect. \ref{sec:reference_data}. The binned data were fitted with a cubic polynomial of the form:
\begin{align}
S\left(T_{\rm eff}\right) = a \cdot \left( \frac{T_{\rm eff} - 4400\,K}{1000\,K} \right)^3 + b.\label{eq:scale_model}
\end{align}
The best fit is shown in Fig. \ref{img:scale_Teff} and the parameters are given in Table \ref{tab:exp_fit_coeff}. Table \ref{tab:cheatsheet} lists the fitted scale factors for specific spectral subtypes and corresponding blueshift velocities.\par
It must be noted that the HARPS sample, despite the calibration of the third-signature model with the values determined in Sect. \ref{sec:reference_data}, still shows a $\sim$6\% smaller CBS at solar temperatures in the fit ($S\left(5800K\right)=0.94$) than expected. This matches the results from \citetads{2017A&A...597A..52M}, who also see a roughly 6\% difference. A comparison to multiple months of HARPS-N Sun-as-a-Star observations \citepads{2021A&A...648A.103D} also show an average scale factor $\sim$5\% below unity after calibration. Section \ref{subsec:activity_influence} demonstrates that deviations of 10\% can be reached by activity influence of sufficient strength and the scatter in the HARPS-N solar observations includes unity within one standard deviation. For this reason it is likely that the FTS spectrum that was used in this work corresponds to a time of higher observable CBS, explaining the Sun-as-a-Star results showing an average blueshift smaller than that, as opposed to an error in our measurements. This is compounded by higher activity in our HARPS sample compared to the Sun that depresses the average CBS values at solar temperatures. As the breadth of scale factors visible at solar temperatures in the HARPS data is mirrored in solar observations, this seems to further indicate an intrinsic noise floor for CBS determinations (see Sect. \ref{subsec:activity_influence} for details).\par
Previous studies into CBS have not been able to find the plateau spanning roughly from 4100\,K to 4700\,K. This is due to them having no stars in this temperature range \citepads{2017A&A...607A.124M}, having low, single-digit numbers \citepads{2009ApJ...697.1032G}, or, with synthetic data, only examining a very narrow slice in temperature \citepads{2018A&A...611A..11C}. As the first study with the required number of stars to cover that range available, we were able to find and report the plateau for the first time. Similarly, the step at 4000\,K could not have been found by previous studies for the same reasons.

\subsection{M stars}
\label{subsubsec:M_results}
Only one M star is shown in Fig. \ref{img:third_sig_examples}, which was of very early M type and shows a scale factor slightly above, but consistent with, zero. Looking at all the M-stars available in Fig. \ref{img:scale_Teff} suggests a CBS of zero, or close to it, in M stars starting at a temperature of 4000\,K. This disagrees with the hypothesis from, among others, \citetads{2003A&A...403.1077K} for Barnard's star. They suggest an increase in CBS with increasing plage area, indicating the underlying convective pattern that is magnetically suppressed is producing convective redshift. They provide a possible explanation, based on M-dwarf convection models from \citetads{2002A&A...395...99L}, in that the decrease in contrast between granules and lanes, combined with a change in vertical flow velocity could lead to a sign change in the balance between granular and intergranular line profile contribution. Our sample includes Barnard's star, though it was initially rejected for unknown $v \sin i$, and its measurements show a scale factor of -0.15$\pm$0.08, matching the approximation from \citetads{2003A&A...403.1077K}, where the convective redshift would be on the order of 33\,m\,s$^{-1}$, about -0.1 in scale factor. This is, however, on the order of the scatter in our results for earlier M stars, while the scatter for mid-M type could not be characterized due to a lack of usable stars. This uncertainty in the actual strength of convective shift is also seen in findings from \citetads{2020A&A...641A..69B}, where they measure the CBS of the M dwarf YZ CMi using the chromatic index and RV variations over one rotation period to model spot and plage distributions. They find a convective redshift on the order of 7 to 237\,m\,s$^{-1}$ or -0.02 to -0.79 in scale factor, encompassing the supposed strength for Barnard's star.\par
A limitation of our results for M dwarfs is that they show comparatively low S/N, as previously shown in Fig. \ref{img:data_S/N_hist}, decreasing scale factor accuracy, and show similar absolute scatter at smaller scale factors compared to solar-like stars at lower scale factors. The general trend does not support the redshift hypothesis and remains consistent with zero scale factor, an extension of the K-dwarf plateau after a relatively large step at 4100\,K. Furthermore, including the reduced $\chi^2$ of the third-signature fit as a size scale for the data points results in Fig. \ref{img:apdx_chi2scale_full}. Larger points here represent better fits and the points are color coded by S/N. The stars not disqualified for one of the aforementioned reasons still show the same behavior, consistent with zero scale factor for M-dwarfs. Looking at the earlier star sample similarly does not show significant changes, with the majority of stars indicating a good fit except toward the edges of the range, the hotter of which was showing higher scatter from the start.\par
In conclusion, our data suggests there to be no general switch to convective redshift for M dwarfs above 3600\,K and generally no convective shift with $\lvert S\rvert > 0.1$. Convective blue- or redshifts above that mark would have been detected. We conclude that there is neither a red- or blueshift at temperatures cooler than 4100\,K, roughly matching the K-to-M-dwarf transition.

\begin{table}
\centering
\caption{Coefficients for the $S\left(T_{\rm eff}\right)$ polynomial fit to the solar-relative scale factors over effective temperature, following Eq. \ref{eq:scale_model}.}
\label{tab:exp_fit_coeff}
\begin{tabular}{c c c c}
\hline\hline
List & $a$ & $b$\\
\hline
all lines & 0.265 & 0.254\\
refined lines & 0.258 & 0.233\\
4500\,K lines & 0.264 & 0.254\\
5500\,K lines & 0.259 & 0.260\\
6000\,K lines & 0.261 & 0.264\\
\hline
\end{tabular}
\tablefoot{"All lines" refers to the full list of lines as taken from VALD, only cleaned of too shallow and too close entries (first part of Sect. \ref{subsec:line_select}). The "refined lines" list is cleaned of those lines that do not give consistently good results (second part of Sect. \ref{subsec:line_select}). The other three are subsets of lines that are also included in cleaned VALD list for stars of the corresponding effective temperature.}
\end{table}

\section{Discussion}
\label{sec:discussion}
We developed a new technique to measure CBS that uses an empirical, ultra-high-resolution solar template of the third signature of granulation that was scaled to fit measurements from HARPS spectra of over 800 usable stars spanning from F to M spectral type. This scale factor is a robust representation of convection strength relative to the Sun and our results show a clear dependence on effective temperature.\par
In this section we discuss the effects of our choice of VALD extraction parameters, especially the temperature, the possible influence of stellar activity, and the lines lower excitation potential. A deeper look into the technical performance and limits of the algorithm is given in Appendix \ref{sec:apdx_algo_perf}.

\subsection{Line list effective temperature selection}
\label{subsec:Teff_effect}
To understand the influence of a 3700\,K based line list, additional lists were extracted from the VALD database (see Table \ref{tab:vald_params}) for 4500\,K, 5500\,K, and 6000\,K. To reduce computation time, the new lists, after a first vetting following Sect. \ref{subsec:line_select} without the post-processing from Appendix \ref{sec:apdx_linerefine}, were checked against the original post-processed 3700\,K one and only the reoccurring lines used each time. The fitted parameters for the polynomial function (Eq. \ref{eq:scale_model}) are given in Table \ref{tab:exp_fit_coeff}. For each line list shown, the scale factor versus effective temperature relations are indistinguishable. Therefore, the choice of effective temperature for the line list does not have a significant impact. The same analysis was carried out for the solar data to check variations in the initial template with the same result.

\subsection{Influence of stellar activity}
\label{subsec:activity_influence}
\begin{figure}
\resizebox{\hsize}{!}{\includegraphics{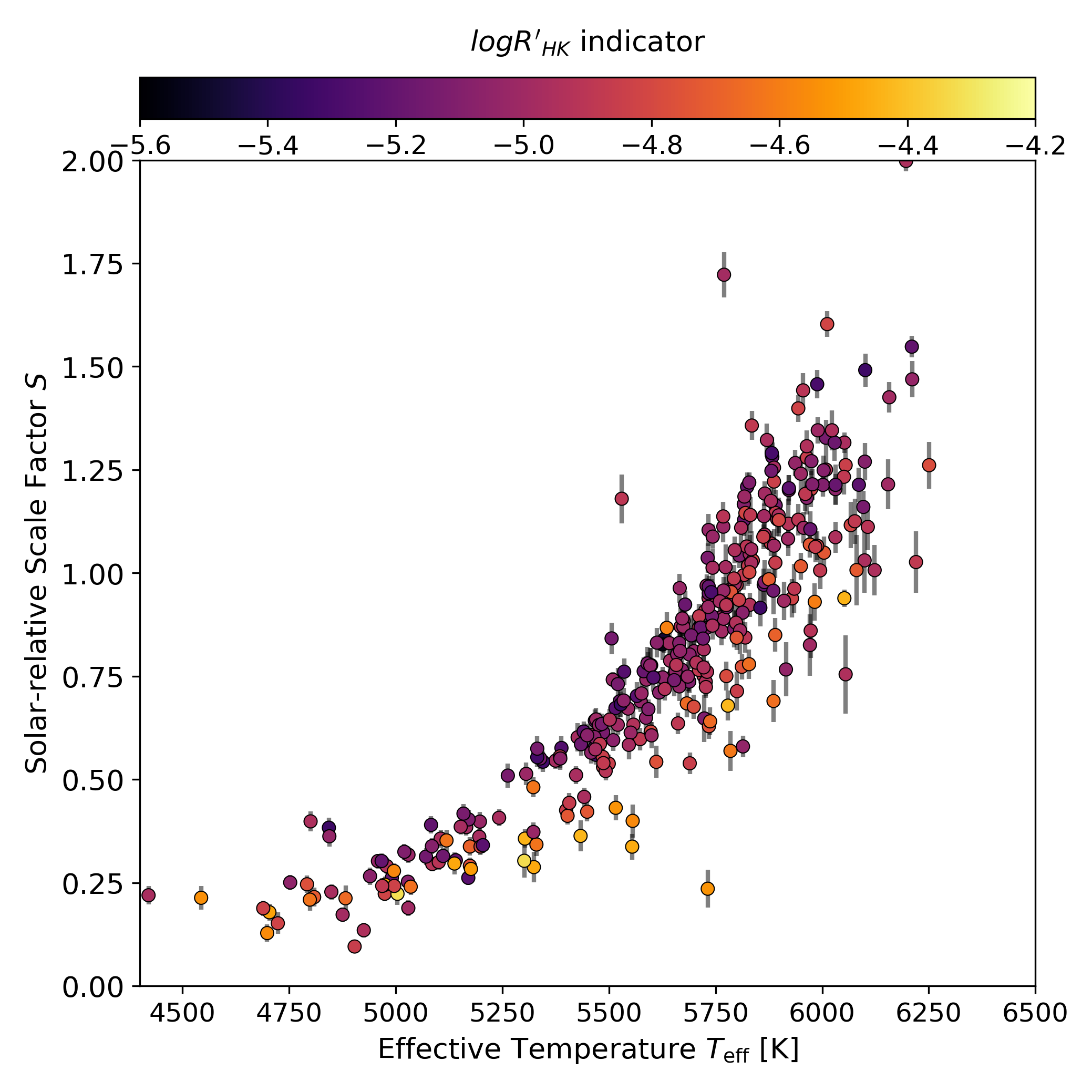}}
\caption{Similar to Fig. \ref{img:scale_Teff} but color-coded with the activity indicator $\log R'_{HK}$. No relation between scatter and activity is readily apparent, but more active stars tend toward lower scale factors.}
\label{img:scale_teff_rhk}
\end{figure}

\begin{figure}
\resizebox{\hsize}{!}{\includegraphics{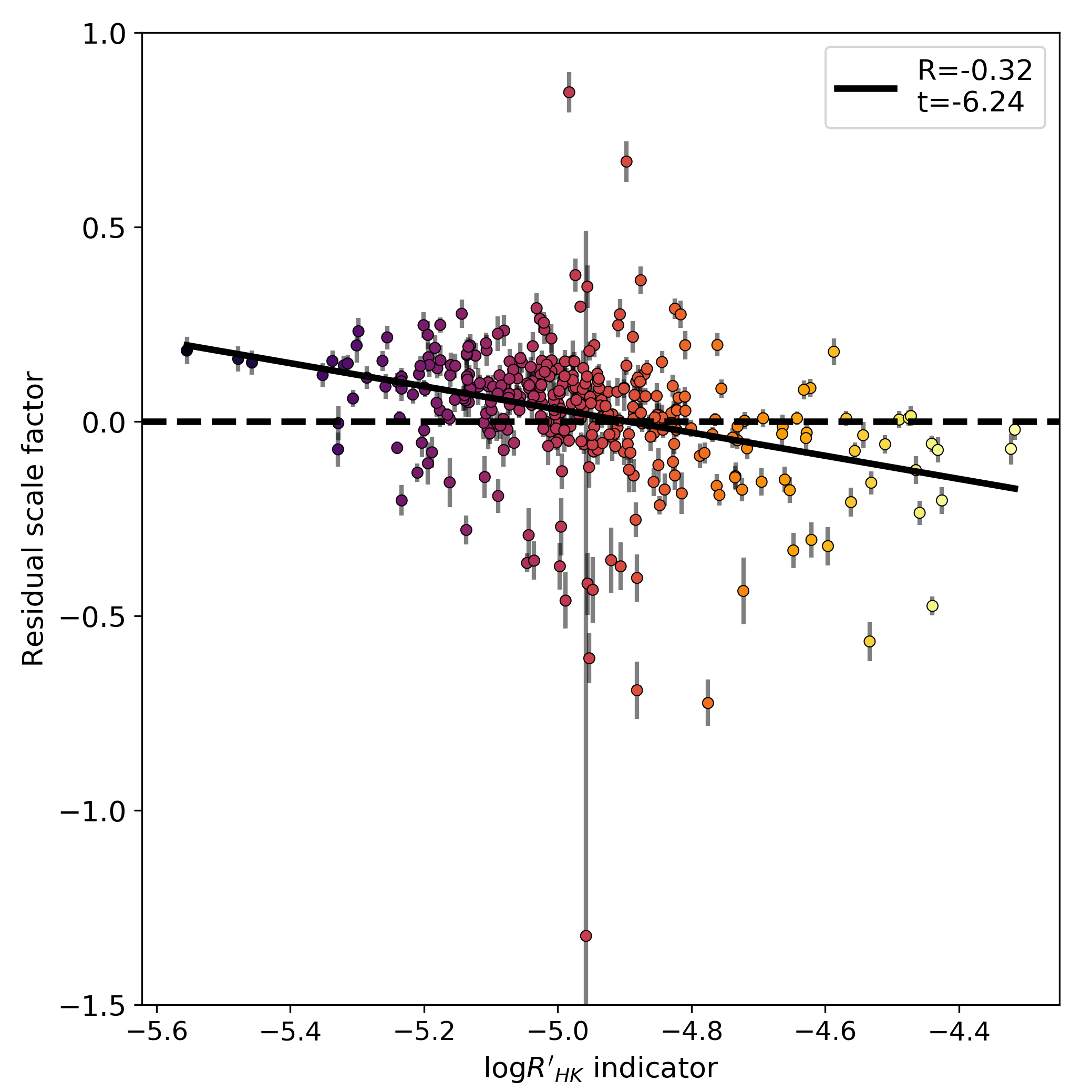}}
\caption{Linear fit (black line) between signed residuals of the scale factor fit and the activity indicator (blue markers). Pearson r and Student t values are shown in the top right.}
\label{img:scale_resid}
\end{figure}

The scatter in the resulting scale factors, as well as the solar templates' deviation from similar temperature HARPS stars as well as HARPS-N solar observations, may be due to different levels of activity between stars of similar spectral type or observation times. This can dampen or enhance CBS or higher activity levels could lead to an increase in noise levels, degrading CBS fit accuracy. To quantify this, we used the $\log R'_{HK}$ indicator values calculated by \citetads{2018A&A...616A.108B} and \citet{marvin_activity}, where we obtained $\log R'_{HK}$ values for 350 stars. In the case of stars with multiple values, we averaged the values. We further provide rotation periods where possible from \citetads{2011arXiv1107.5325L} in Table \ref{tab:star_data}. The results are color-coded for activity in Fig. \ref{img:scale_teff_rhk} and listed in Table \ref{tab:star_data}. A slight trend toward lower scale factors for higher levels of activity is visible. The absolute residual of the polynomial fit correlates with the activity indicator with a Pearson correlation coefficient of $r$=0.02 with a t-Value of 0.39 (p-Value $\sim$0.7), indicating higher activity does not lead to higher scatter. Figure \ref{img:scale_resid} shows the signed residual, which has correlation coefficients of $r$=-0.32 and $t$=-6.24. From this we conclude that higher activity inhibits CBS strength (p-Value < $10^{-5}$). This supports the findings from \citetads{2017A&A...597A..52M} and \citetads{2017A&A...607A.124M} who published an anticorrelation between CBS and activity. They further find a link from CBS to metallicity, which seems to provide a competing effect with higher metallicities correlating with smaller CBS at similar activity levels. The metallicity-activity dependence however goes the opposite: high metallicities correlate with lower activity levels, which correlate to higher CBS strengths. The reason for this discrepancy is unknown, though they postulate it might either be due to metallicity changes contaminating the $\log R'_{HK}$ determination or physical changes in the small-scale convection pattern. Inhibition of CBS through higher stellar activity is better understood (see e.g., \citetads{2018A&A...610A..52B}), as magnetic pressure in active regions counteracts convective motion and lowers the net blueshift.\par
These results do not explain the solar template deviation however. The solar FTS was recorded in 2014, before the HARPS-N observations and, according to sunspot numbers, during a time of higher solar activity \citep{sidc}\footnote{\url{http://www.sidc.be/silso/}}. This should correspond to a lower level of CBS, the opposite to the observation. As solar activity in terms of $\log R'_{HK}$ varies very little (HARPS-N solar observations cover -4.96 to -5.04, about the average of the HARPS sample) this translates into negligible CBS suppression, eliminating activity as the source of the discrepancy. Coupled with the HARPS-N observations including unity within one standard deviation and showing strong dispersion even between observations taken on the same day, this indicates a CBS noise floor of $\sim$10\% for the Sun, irrespective of activity, and operating on the granulation timescale of $\sim$10 minutes. This is in accordance with \citetads{2019MNRAS.487.1082C}, who also see intraday variations in the CCF BIS on the 10\% level.

\subsection{Influence of excitation potential}
\citetads{1981A&A....96..345D} reported a dependence of CBS on the lower excitation potential of the spectral lines. They postulated that higher temperatures in granules produce stronger excited lines, which then show a correspondingly higher blueshift compared to lower excited lines from intergranular lanes. In their Fig. 3 they showed a plot of excitation potentials for lines of intermediate depth against blueshift, which showed a highly scattered dependence. To investigate if this has any impact on the results obtained in this analysis, we repeated their analysis by restricting our results to the same sample of \ion{Fe}{i} lines on the solar FTS spectrum. We did not find any relation between CBS and excitation potential in our data, matching nonfindings by \citetads{2012AJ....143...92G} for giants and supergiants.

\subsection{Comparison to other works}
\label{subsec:Meunier_comp}
\begin{figure}
\resizebox{\hsize}{!}{\includegraphics{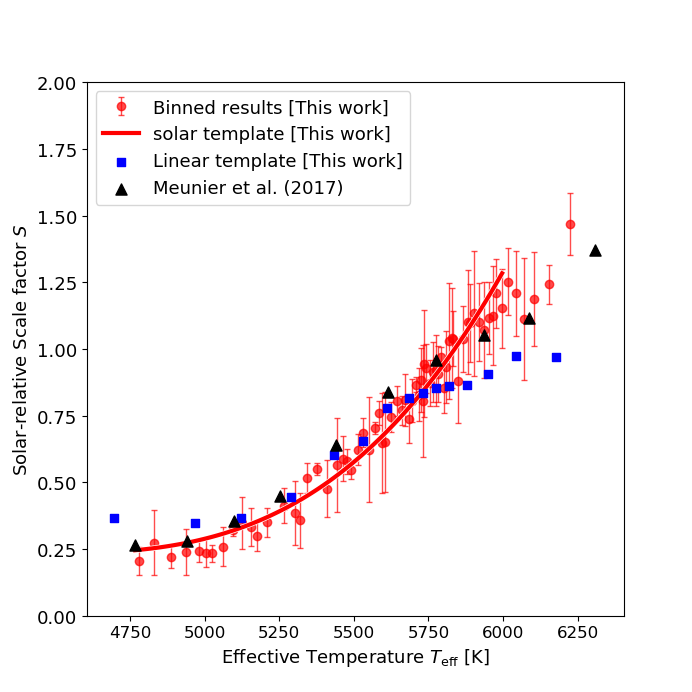}}
\caption{Comparison of the results from \citetads{2017A&A...607A.124M} (black triangles) with this work (red, solid line and red markers with error bars) and an alternate, linear template (blue squares).}
\label{img:Meunier_compare}
\end{figure}
\citetads{2017A&A...597A..52M} and \citetads{2017A&A...607A.124M} utilized a simpler approach to measure the strength of CBS in the form of a linear model $v_{\rm conv}\left(d\right)$ for the third-signature fit in place of a cubic polynomial scaled from a high-quality solar template (Eq. \ref{eq:template_third_sig}). It was carried out on a similar albeit smaller sample of HARPS observations that spans only 167 and 360 stars, respectively, and extends down to a temperature of 5000\,K (4700\,K), excluding the later K dwarfs present in this work. They also find a decrease in CBS, proxied in their work by the "Third Signature Slope" (TSS), with decreasing effective temperature, though their trend, as shown in Fig. \ref{img:Meunier_compare}, recreated from Fig. 5 in \citetads{2017A&A...607A.124M} and normalized to solar values, appears linear unlike what our results based on Eqs. \ref{eq:scale_model} and \ref{eq:template_third_sig} suggest. Overlaying our results reveals a good, though slightly offset, agreement for stars at 5900\,K and cooler but deviates strongly for stars above that. This may be explained by the fact that these weaker third signatures (scale factors < 1.0) have a smaller curvature and could be better approximated as linear without significant loss of precision. This no longer holds for stronger third signatures (scale factor > 1.0), where the curvature becomes stronger. In this case it is to be expected that a linear model underestimates the strength of the signature, especially since the shallower lines are still comparatively linear with a smaller slope and make up approximately 2/3 of the depth range before the signature curves up significantly. This is compounded by them only choosing a range of depths 0.4 to 0.8, excluding the deeper, much more curved, part of the third signature we have included in our model (up to depth 0.95). Switching our solar template for a linear model and restricting the fit to their depth range results in a qualitatively similar relation, though the very slight sigmoid shape inherent in their results appears much more pronounced, approaching the form of a logistic function. Differences exist for hotter and cooler stars, appearing as if our TSS results are compressed, and in our solar TSS (TSS$_{\odot}$), the linear slope of the template, which shows as 575\,m\,s$^{-1}$ instead of 776\,m\,s$^{-1}$. A potential influence is the choice of line list, which can lead to significant differences if wide applicability is not kept in mind, similar to our rejection of the \citetads{1994ApJS...94..221N} line list, which is over-adapted to the Sun. Another likely contributing factor to the deviations are the significantly changing ionization fractions of iron at temperatures above 6000\,K. The ionization balance between \ion{Fe}{i} and \ion{Fe}{ii} reverses around 6000 to 7000\,K, reducing the vertical atmospheric range in which \ion{Fe}{i} lines can form and therefore the contributing velocity components. Limiting those components to the cooler, upper parts of the atmosphere restricts the line contribution to lower velocities and one would expect a smaller observed CBS compared to lines that have a higher temperature flipping point in their species ionization fractions. Qualitatively, the choice between solar and linear template seems to be the cause of the difference in the hotter stars scale factors.\par
Our results are in agreement with the theoretical models of \citetads{2014arXiv1405.7628M} who predict an increase in convection velocity (and therefore CBS strength over all line depths) and granule brightness, hence a stronger CBS signature, for either hotter or lower surface gravity stars or both, which our results fully support. The increase in scale factor matches the theoretically expected small increase in granule size and significant increase in granule contrast due to increasing effective temperatures compounded by the contribution from the slightly decreasing surface gravity.\par
We also compare our results to the theoretical results of \citetads{2018A&A...611A..11C}. Unlike \citetads{2017A&A...597A..52M} they used purely synthetic spectra to measure CBS with cross-correlation techniques for a wide range of effective temperatures. Their resulting CBS also increases with effective temperature; however, their results appear to underestimate CBS on the parameter range above 5400\,K by at least 100\,m\,s$^{-1}$ compared to our CBS values. The results roughly match below 5000\,K, but are hindered by dispersion on the order of 200\,m\,s$^{-1}$ and discontinuous coverage. The data are instead clustered in narrow temperature slices. Accounting for those deviations, which they point out themselves, puts their results much closer to ours. The last major deviation is their predicted flattening of CBS for stars above 6000\,K. This is potentially compatible with our results since our binned results show the beginnings of a plateau. A definite conclusion is hindered by the lack of usable stars hotter than 6200\,K in our sample. 
As for \citetads{2017A&A...597A..52M}, it is also important to remember that the choice of spectral lines can have a large impact if generality is lost. Unlike in our work, \citetads{2018A&A...611A..11C} are using only \ion{Fe}{i} lines and may be susceptible to the mentioned changes in ionization fractions. Therefore, comparisons of results above 6000\,K, namely the plateau in the \citetads{2018A&A...611A..11C} data, are not applicable unless one were to also use only \ion{Fe}{i} lines, which would not be representative for the stars' actual CBS. \citetads{2017A&A...597A..52M}, while not stating the actual lines they used, have based their list on a catalog of \ion{Fe}{ii}, on top of \ion{Fe}{i}, taking care of that problem, while our list also includes some \ion{Fe}{ii} lines.\par
\citet{basu2017} present a mathematical approach to model observable CBS velocity amplitudes based on approximations for granule coverage and sizes using mixing-length theory and established scaling relations. We applied their results as a scaling relation and inserted stellar values from our sample, which roughly matches our results in general shape. Depending on where in our sample we calibrated the scaling origin to, it either matches the range 4700-5500\,K or the 5400-6000\,K, though not both. It also does not reproduce the K-dwarf plateau in either case. A likely cause is their assumption that individual granules contribute the same intensity fluctuations irrespective of spectral type and that the cancellation between granular blueshift and intergranular redshift is also independent of spectral type, which they call rather simplistic themselves. The scaling does predict a near-zero CBS starting at early M-stars for decreasing effective temperature, which is mirrored in our results.

\subsection{Extending the data set}
We show in Sect. \ref{subsubsec:M_results} that our results on the convective shift of M dwarfs indicate zero CBS. We could not supplement our HARPS data using high-resolution, higher-S/N observations from the dedicated M-dwarf survey CARMENES \citepads{2016SPIE.9908E..12Q}, since the instrument does not cover the shorter wavelength range between 450 and 500\,nm that a large part of the spectral lines used in this work fall into and the expanded red end compared to HARPS does not contain enough usable atomic lines to compensate. A revisit once more HARPS observations are available, increasing coadded S/N especially for later M-dwarfs, is expected to improve the situation.\par
Additional extension possibilities include stars above $\sim$6000\,K to observe convective shift behavior close to and above the granulation boundary, investigating claims by \citetads{2018A&A...611A..11C} of a plateau in CBS as well as the influence of changes in the ionization fractions. This would need either a sample of low $v \sin i$ F stars or a revision of the line-core fitting procedure.\par
Similarly, we are working on an investigation of a set of subgiant and giant spectra, extending the work of \citetads{2009ApJ...697.1032G} toward understanding the CBS behavior of stars that left the main sequence.

\section{Conclusion}
\label{sec:conclusion}
Our new approach to measure CBS that used a ultra-high-resolution solar template applied to high-resolution stellar measurements, has proven to be robust over a large range of effective temperatures, from early-M to early-G or late-F spectral type or 3800\,K to 6000\,K. Further, we were able to determine limits in terms of S/N and minimal resolving power required to apply our technique, both of which are very reasonable in terms of instrumental requirements.\par
We used the template and a large sample of coadded HARPS main sequence star spectra to obtain the following results:
\begin{itemize}
\item We provide a revised model for the solar third signature of granulation based on an expanded list of lines that is also applicable to other stars, unlike previously used lists that were curated for the Sun.
\item We confirm that CBS strength scales strongly with effective temperature above 4700\,K and provide a fitted scaling relation. Unlike previous findings, our results scale with the third power of effective temperature. The discrepancy appears to be due to differences between our empirical and previous simplified third-signature models. They agree within the margins expected due to different line lists and said algorithmic differences
\item We find that between 4100\,K and 4700\,K CBS remains constant at 23\% solar strength. Previous studies do not sufficiently cover this region to make this determination.
\item Between 4100\,K and 3800\,K, we find that CBS shows a sharp step down and appears to remain constant, though scatterd, around zero observable CBS. The topic of CBS in M dwarfs has been highly debated and could not be definitely resolved.
\item The preceding results prove the applicability of the third-signature fitting approach, shown by \citetads{2012AJ....143...92G} for giants, to main sequence stars, as indicated by \citetads{2009ApJ...697.1032G}.
\item We confirm that stellar activity correlates with lower CBS, explaining part of the dispersion in our results.
\item We could not find any dependence of CBS on a spectral lines lower excitation potential.
\item We provide an expanded third-signature model for the Sun, taking into account wavelength in addition to line depth, demonstrating the strong effect the former has on CBS.
\item We provide synthetic and empirically derived limits on our approach to determine CBS strengths as well as corrections to account for the effect of instrumental resolving power.
\end{itemize}
Further, the technique and its much more reasonable requirements on data quality compared to the classical bisector approach opens up the study of CBS to a significantly increased number of instruments and research groups. It also allows fainter stars, be it from distance or temperature, to be studied without the need to coadd dozens of spectra, reducing the required observation time. The technique also standardizes the way one may talk about CBS strength as the scale factor is agnostic to any otherwise arbitrary choice in reference line, its central wavelength, and absorption depth.

\begin{acknowledgements}
We thank the referee for their insightful report and for their comments that have improved the clarity of the paper. FL acknowledges the support of the DFG priority program SPP 1992 "Exploring the Diversity of Extrasolar Planets (RE 1664/18). SVJ acknowledges the support of the German Science Foundation (DFG) Research  Unit  FOR2544  `Blue  Planets  around  Red  Stars',  project  JE  701/3-1  and  DFG priority program SPP 1992 `Exploring the Diversity of Extrasolar Planets' (JE 701/5-1). The SIMBAD database\footnote{\url{http://simbad.u-strasbg.fr/simbad/}}, hosted at the CDS, Strasbourg, France, was used in this research. This research has made use of NASA's Astrophysics Data System Bibliographic Services\footnote{\url{http://adsabs.harvard.edu/}}. This work has made use of the VALD database\footnote{\url{http://vald.astro.uu.se/}}, operated at Uppsala University, the Institute of Astronomy RAS in Moscow, and the University of Vienna. This work has made use of data from the European Space Agency (ESA) mission {\it Gaia} (\url{https://www.cosmos.esa.int/gaia}), processed by the {\it Gaia} Data Processing and Analysis Consortium (DPAC, \url{https://www.cosmos.esa.int/web/gaia/dpac/consortium}). Funding for the DPAC has been provided by national institutions, in particular the institutions participating in the {\it Gaia} Multilateral Agreement. We thank Trifon Trifonov for providing the coadded spectra they created to be used in this work. The original observations were collected at the European Organization for Astronomical Research in the Southern Hemisphere under ESO programmes: 0100.C-0097, 0100.C-0111, 0100.C-0414,0100.C-0474, 0100.C-0487, 0100.C-0750, 0100.C-0808, 0100.C-0836, 0100.C-0847,  0100.C-0884,  0100.C-0888,  0100.D-0444,  0100.D-0717,  0101.C-0232,0101.C-0274, 0101.C-0275, 0101.C-0379, 0101.C-0407, 0101.C-0516, 0101.C-0829,  0101.D-0717,  0102.C-0338,  0102.D-0717,  0103.C-0548,  0103.D-0717,060.A-9036,  060.A-9700,  072.C-0096,  072.C-0388,  072.C-0488,  072.C-0513,072.C-0636, 072.D-0286, 072.D-0419, 072.D-0707, 073.A-0041, 073.C-0733,073.C-0784, 073.D-0038, 073.D-0136, 073.D-0527, 073.D-0578, 073.D-0590,074.C-0012,  074.C-0037,  074.C-0102,  074.C-0364,  074.D-0131,  074.D-0380,075.C-0140,  075.C-0202,  075.C-0234,  075.C-0332,  075.C-0689,  075.C-0710,075.D-0194, 075.D-0600, 075.D-0614, 075.D-0760, 075.D-0800, 076.C-0010,076.C-0073,  076.C-0155,  076.C-0279,  076.C-0429,  076.C-0878,  076.D-0103,076.D-0130, 076.D-0158, 076.D-0207, 077.C-0012, 077.C-0080, 077.C-0101,077.C-0295, 077.C-0364, 077.C-0530, 077.D-0085, 077.D-0498, 077.D-0633,077.D-0720,  078.C-0037,  078.C-0044,  078.C-0133,  078.C-0209,  078.C-0233,078.C-0403, 078.C-0751, 078.C-0833, 078.D-0067, 078.D-0071, 078.D-0245,078.D-0299,  078.D-0492,  079.C-0046,  079.C-0127,  079.C-0170,  079.C-0329,079.C-0463,  079.C-0488,  079.C-0657,  079.C-0681,  079.C-0828,  079.C-0927,079.D-0009, 079.D-0075, 079.D-0118, 079.D-0160, 079.D-0462, 079.D-0466,080.C-0032,  080.C-0071,  080.C-0664,  080.C-0712,  080.D-0047,  080.D-0086,080.D-0151, 080.D-0318, 080.D-0347, 080.D-0408, 081.C-0034, 081.C-0119,081.C-0148,  081.C-0211,  081.C-0388,  081.C-0774,  081.C-0779,  081.C-0802,081.C-0842, 081.D-0008, 081.D-0065, 081.D-0109, 081.D-0531, 081.D-0610,081.D-0870,  082.B-0610,  082.C-0040,  082.C-0212,  082.C-0308,  082.C-0312,082.C-0315,  082.C-0333,  082.C-0357,  082.C-0390,  082.C-0412,  082.C-0427,082.C-0608,  082.C-0718,  083.C-0186,  083.C-0413,  083.C-0794,  083.C-1001,  083.D-0668,  084.C-0185,  084.C-0228,  084.C-0229,  084.C-1039,  085.C-0019,085.C-0063,  085.C-0318,  085.C-0393,  086.C-0145,  086.C-0230,  086.C-0284,086.D-0240,  087.C-0012,  087.C-0368,  087.C-0649,  087.C-0831,  087.C-0990,087.D-0511,  088.C-0011,  088.C-0323,  088.C-0353,  088.C-0513,  088.C-0662,089.C-0006,  089.C-0050,  089.C-0151,  089.C-0415,  089.C-0497,  089.C-0732,089.C-0739,  090.C-0395,  090.C-0421,  090.C-0540,  090.C-0849,  091.C-0034,091.C-0184,  091.C-0271,  091.C-0438,  091.C-0456,  091.C-0471,  091.C-0844,091.C-0853,  091.C-0866,  091.C-0936,  091.D-0469,  092.C-0282,  092.C-0454,092.C-0579,  092.C-0721,  092.C-0832,  092.D-0261,  093.C-0062,  093.C-0409,093.C-0417,  093.C-0474,  093.C-0919,  094.C-0090,  094.C-0297,  094.C-0428,094.C-0797,  094.C-0894,  094.C-0901,  094.C-0946,  094.D-0056,  094.D-0596,095.C-0040,  095.C-0105,  095.C-0367,  095.C-0551,  095.C-0718,  095.C-0799,095.C-0947,  095.D-0026,  095.D-0717,  096.C-0053,  096.C-0082,  096.C-0183,096.C-0210,  096.C-0331,  096.C-0417,  096.C-0460,  096.C-0499,  096.C-0657,096.C-0708,  096.C-0762,  096.C-0876,  096.D-0402,  096.D-0717,  097.C-0021,097.C-0090,  097.C-0390,  097.C-0434,  097.C-0561,  097.C-0571,  097.C-0864,097.C-0948,  097.C-1025,  097.D-0156,  097.D-0717,  098.C-0269,  098.C-0292,098.C-0304,  098.C-0366,  098-C-0518,  098.C-0518,  098.C-0739,  098.C-0820,098.C-0860,  098.D-0717,  099.C-0093,  099.C-0138,  099.C-0205,  099.C-0303,099.C-0304,  099.C-0374,  099.C-0458,  099.C-0491,  099.C-0798,  099.C-0880,099.C-0898, 099.D-0717, 1101.C-0721, 180.C-0886, 183.C-0437, 183.C-0972,183.D-0729,  184.C-0639,  184.C-0815,  185.D-0056,  188.C-0265,  188.C-0779,190.C-0027,  191.C-0505,  191.C-0873,  192.C-0224,  192.C-0852,  196.C-0042,196.C-1006,  198.C-0169,  198.C-0836,  198.C-0838,  281.D-5052,  281.D-5053,282.C-5034,  282.C-5036,  282.D-5006,  283.C-5017,  283.C-5022,  288.C-5010,292.C-5004, 295.C-5031, 495.L-0963, 60.A-9036, 60.A-9700, and 63.A-9036.
The analysis was carried out using the programming language Python3\footnote{\url{https://www.python.org/}}, and the accompanying software packages: Numpy\footnote{\url{https://numpy.org/}}, Scipy\footnote{\url{https://www.scipy.org/scipylib/}}, Astropy\footnote{\url{https://www.astropy.org/}}, Astroquery\footnote{\url{https://astroquery.readthedocs.io}}, Matplotlib\footnote{\url{https://matplotlib.org/}}, and PyAstronomy\footnote{\url{https://github.com/sczesla/PyAstronomy}}\citep{pya}.
\end{acknowledgements}

\bibliographystyle{aa} 
\bibliography{paper} 

\begin{appendix}
\section{Line list refinement}
\label{sec:apdx_linerefine}
\begin{figure}
\resizebox{\hsize}{!}{\includegraphics{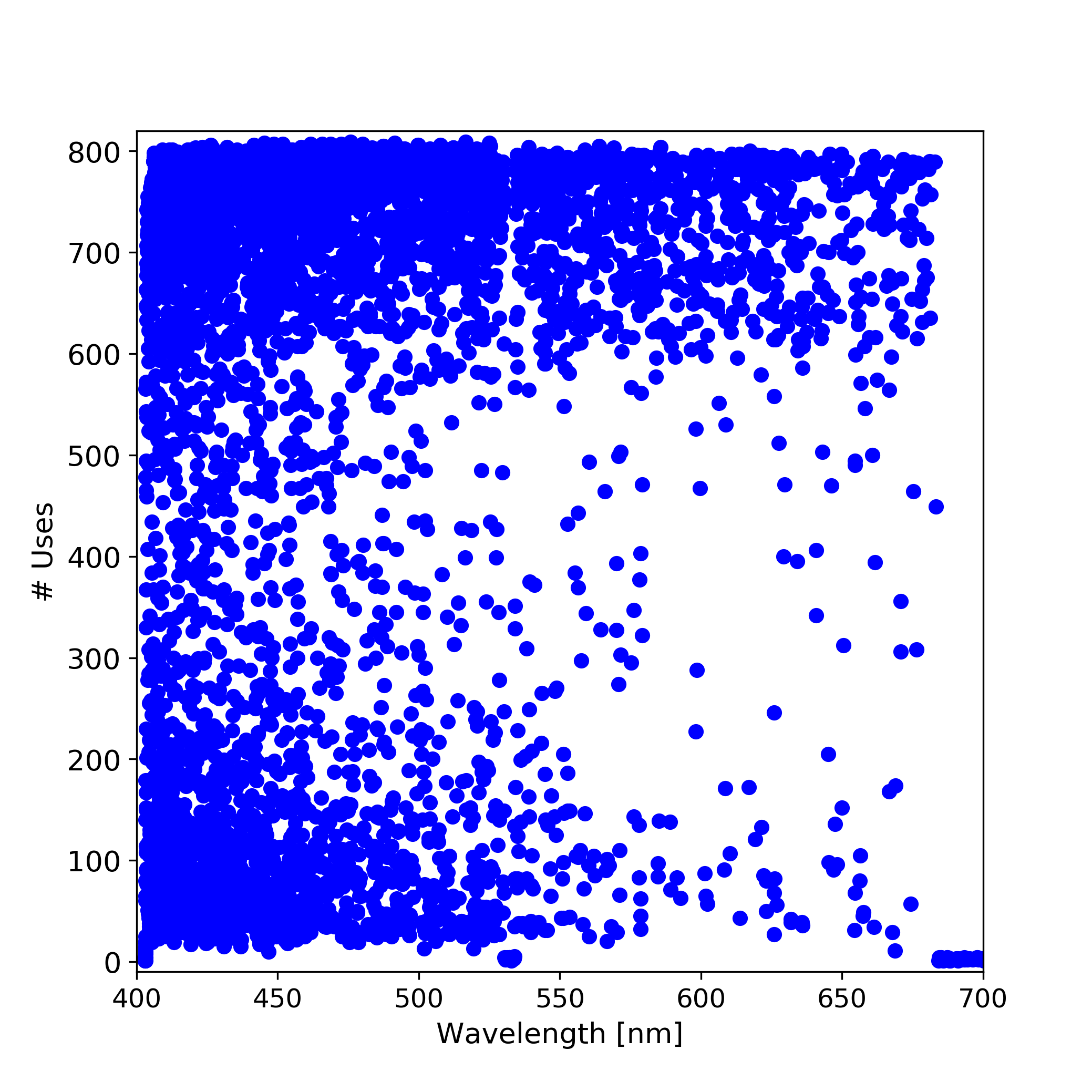}}
\caption{Initial line utilization; number of stars per line over wavelength.}
\label{img:Line_used}
\end{figure}

\begin{figure}
\resizebox{\hsize}{!}{\includegraphics{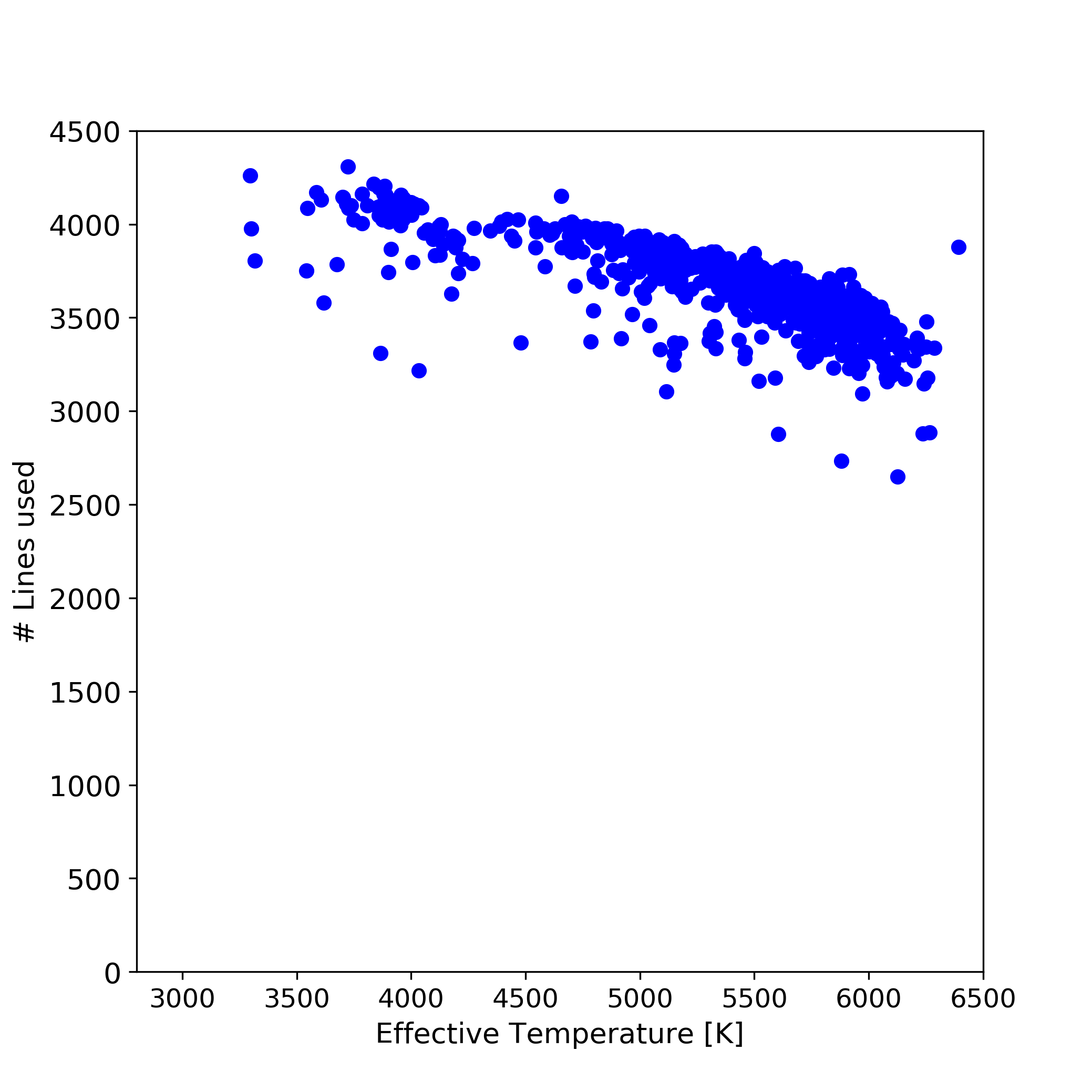}}
\caption{Initial line numbers; number of lines per star over $T_{\rm eff}$.}
\label{img:Line_numbers}
\end{figure}

\begin{figure*}
\resizebox{\hsize}{!}{\includegraphics{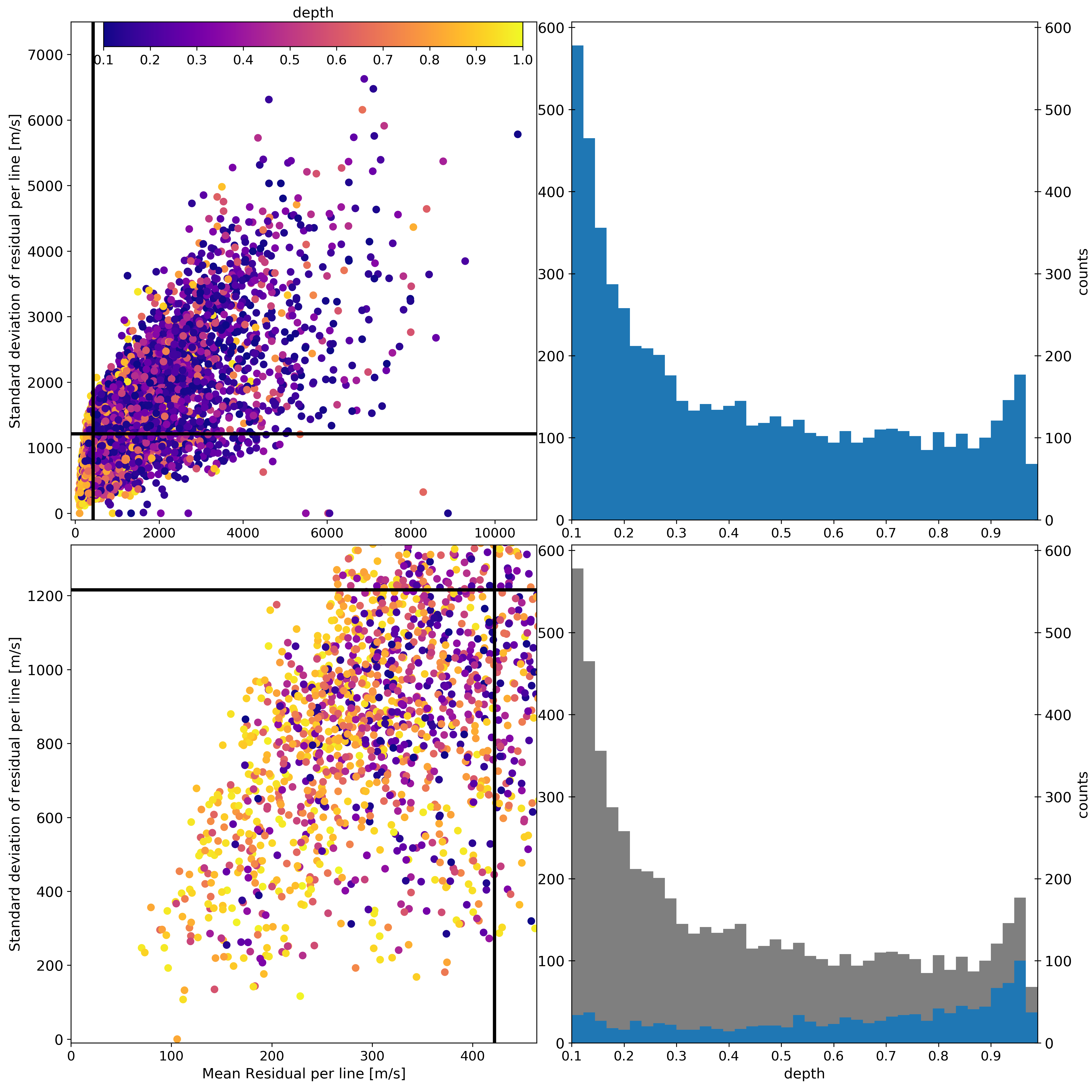}}
\caption{Percentile-based line list vetting. Left panels: Mean and STD of residuals per line over all stars. Marked are 25th percentile for mean and 50th for std. Colors indicate line absorption depth from the VALD 3700\,K list. Right panels: Corresponding histograms of line depth counts for the full set of lines (top) and the percentile filtered subset (bottom).}
\label{img:Line_res_stds}
\end{figure*}

\begin{figure}
\resizebox{\hsize}{!}{\includegraphics{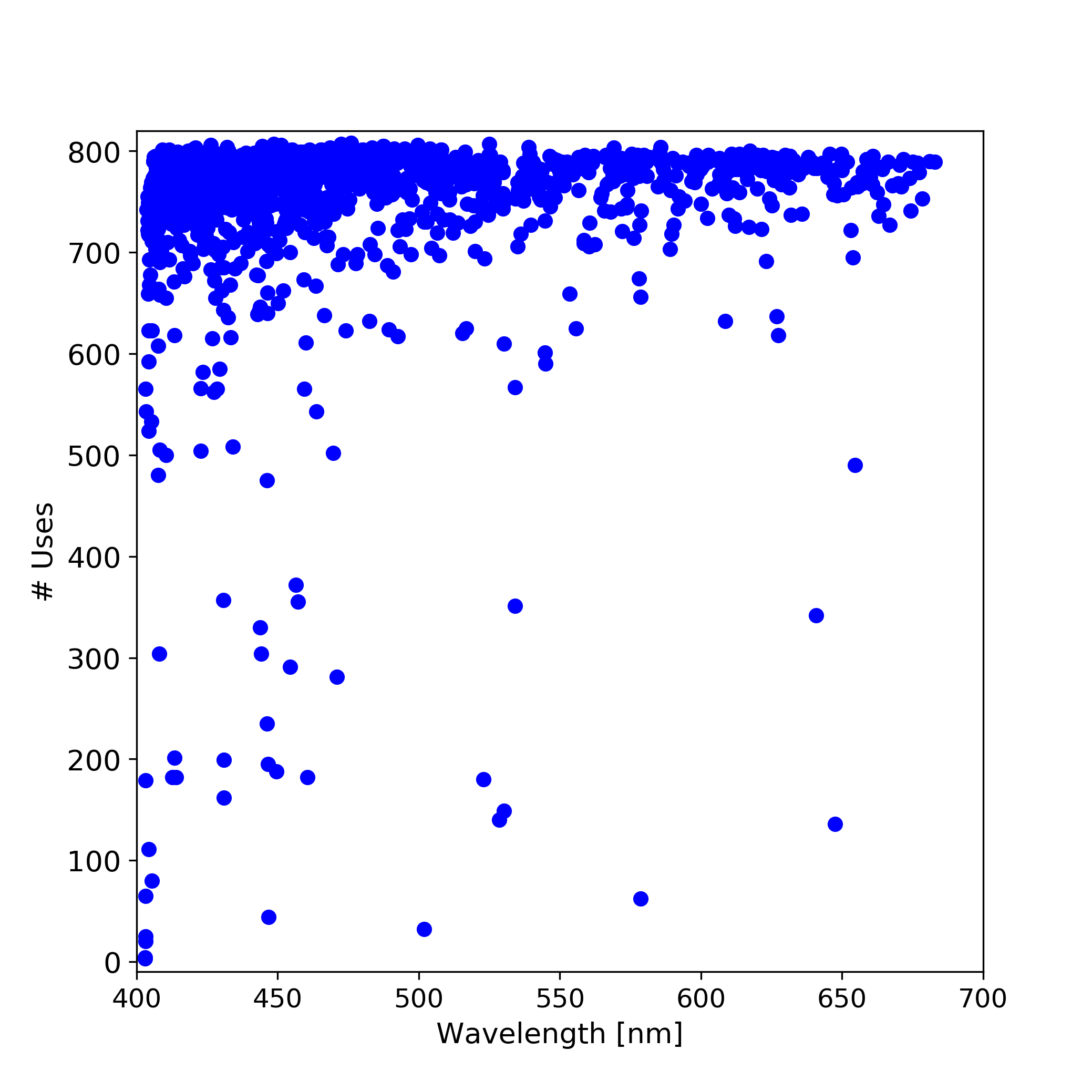}}
\caption{Remaining line utilization; number of stars per line over wavelength.}
\label{img:retained_Line_used}
\end{figure}

\begin{figure}
\resizebox{\hsize}{!}{\includegraphics{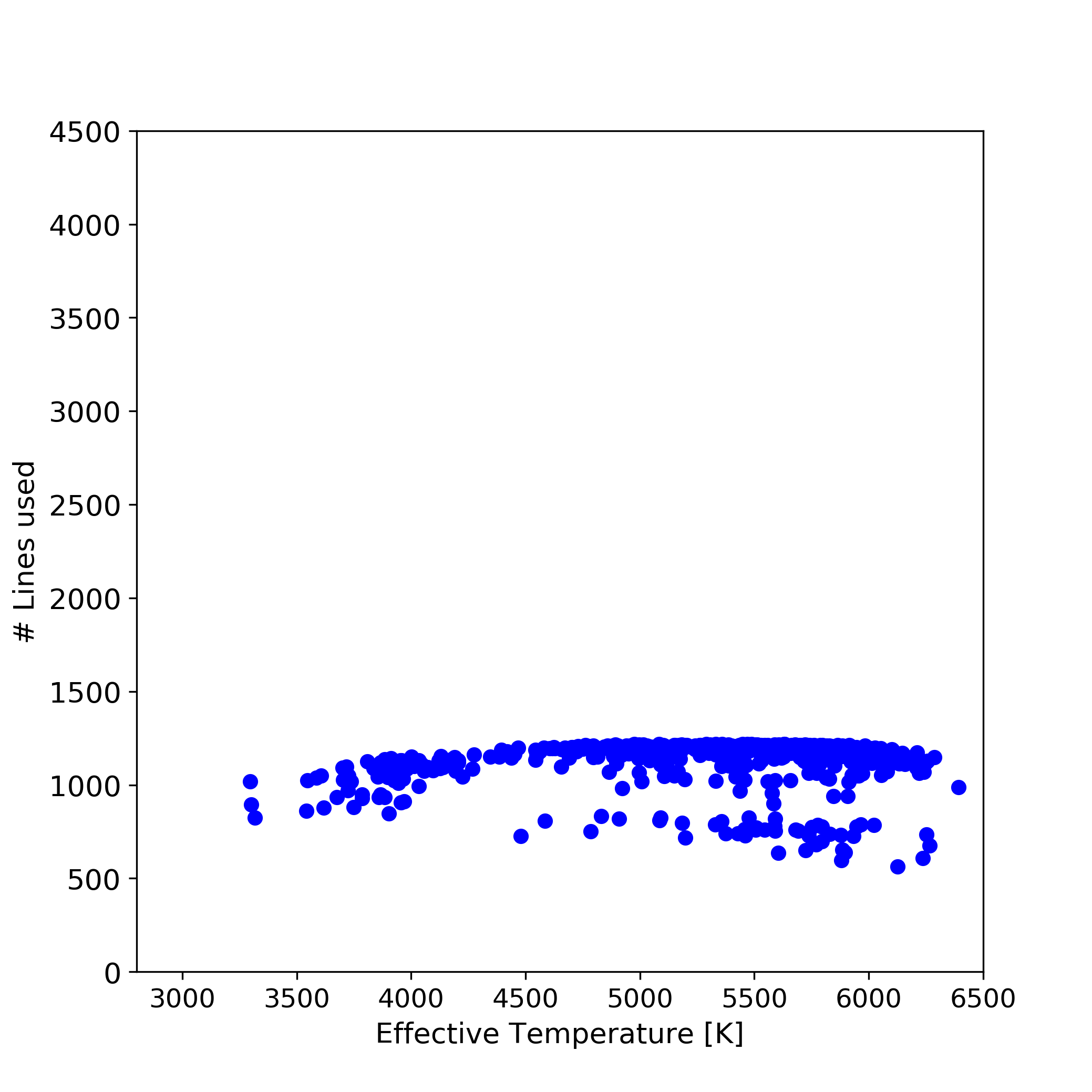}}
\caption{Remaining line numbers; number of lines per star over $T_{\rm eff}$.}
\label{img:retained_Line_numbers}
\end{figure}

In Sect. \ref{subsec:line_select} we explain how we created the list of lines used in this work. An important part was that the base line list obtained from VALD still contains a large number of lines that are present in a star matching the parameters used for the extraction, but not necessarily in all stars in the intended spectral range from late-F dwarfs to M dwarfs. It also does not guarantee that a specific line, while present, is reliably measurable. For this reason the prefiltered line list was post-processed after the first round of measurements to remove such occurrences. The measuring algorithm developed in this work already flags lines that could not be measured accurately. Investigating the line utilization (Fig. \ref{img:Line_used}) it is readily apparent that a good part of them is used only rarely, but also that numerous lines are used in nearly all stars. It can also be seen that the blue end, up to about 500\,nm contains the majority of usable lines, limiting the use of data from instruments such as CARMENES, which are geared more toward red wavelengths. Figure \ref{img:Line_numbers} shows the number of lines used per star. One can see that with 3000-4000 lines per star, there is room for vetting the line list. Figure \ref{img:Line_res_stds} illustrates how the vetting step worked by plotting for each line the standard deviation of the residual between the fitted third signature and the measured line position over all stars against the mean residual. It can be seen that many lines are intrinsically less suited for the third-signature determination, showing large deviations over the entire sample. They were identified and excluded by retaining only lines that had a standard deviation that is not zero, meaning that they are used more than once, whose mean residual fell below the 25th percentile, and which had a standard deviation below the 50th percentile. These values were found as a compromise between number of lines retained and quality of the resulting third signatures. Redoing the first two figures (now Fig. \ref{img:retained_Line_used} and \ref{img:retained_Line_numbers}) all the seldomly used lines are gone without impacting the wavelength coverage. For each star the vetting means a reduction in line use by about a factor of two, leaving FGK stars with about 1300 lines each. M-stars are reduced to about 1000 lines.\par
The removed lines are generally listed as shallow in the VALD 3700\,K list, as can be seen by the color coding in Fig. \ref{img:Line_res_stds}, which is not surprising since they are harder to measure accurately due to their comparatively higher width relative to their depth. This makes them more susceptible to noise. The mean and standard deviation are correlated with depth at Pearson $r$=-0.35 and -0.30. Wavelength shows a much weaker, effectively irrelevant, correlation at $r$=-0.03 and 0.001, respectively. By filtering, the majority of lines involved are no longer located at shallow depths but changed to deep lines. Magnetic sensitivity, that was approximated by the Land\'e factor, seems to play no role in the filtering at R=0.04 and 0.02 for mean and standard deviation, respectively.

\section{Wavelength dependence of CBS}
\label{sec:apdx_lam_effect}
\begin{figure}
\resizebox{\hsize}{!}{\includegraphics{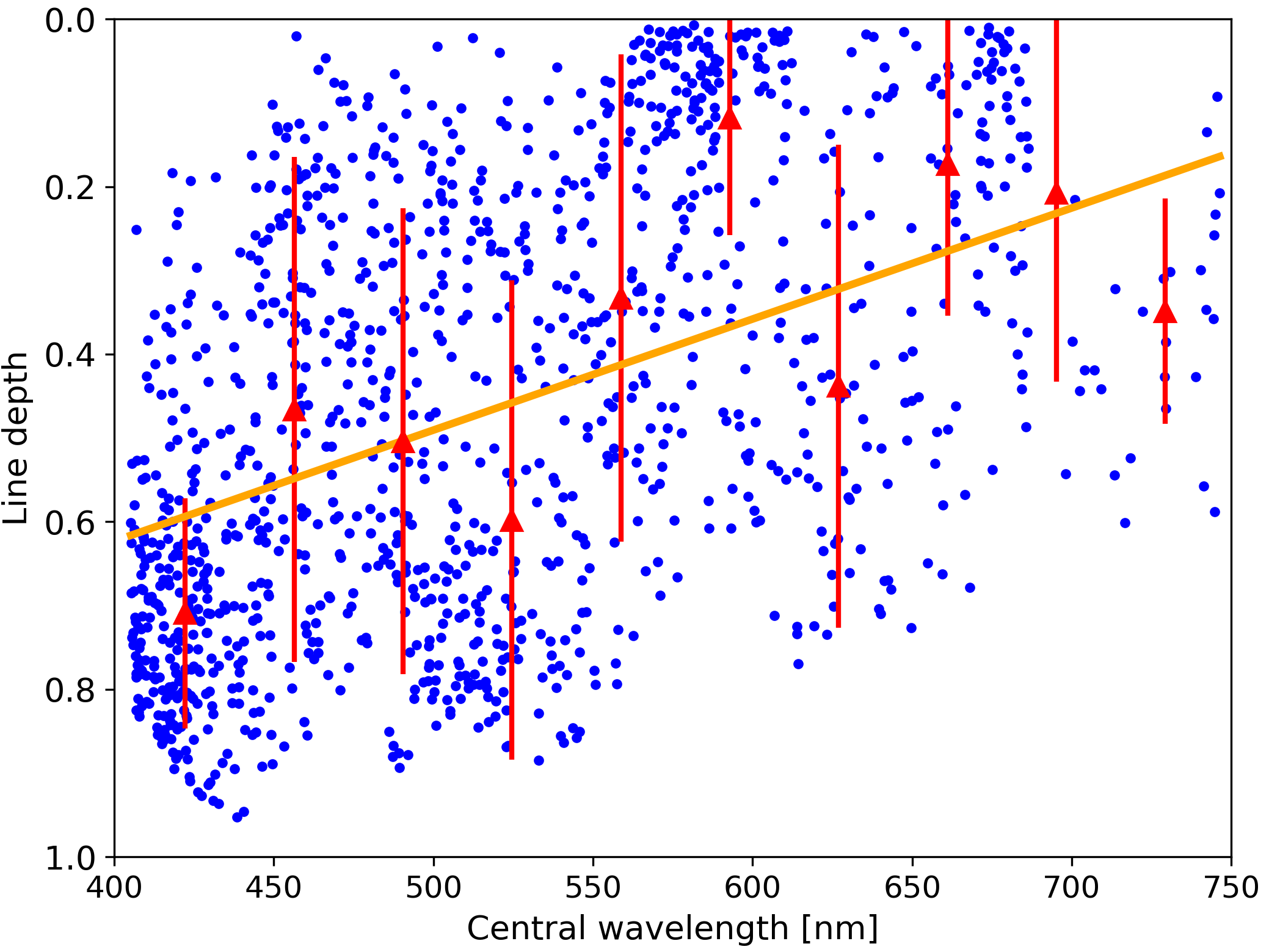}}
\caption{Distribution of line absorption depth with central wavelength (blue dots). The binned data are plotted in red with error bars. A linear fit is shown as an orange curve. The trend toward deep lines in the blue region is readily apparent as is the higher density of lines.}
\label{img:solar_sig_depth_wav}
\end{figure}

\begin{figure*}
\resizebox{\hsize}{!}{\includegraphics{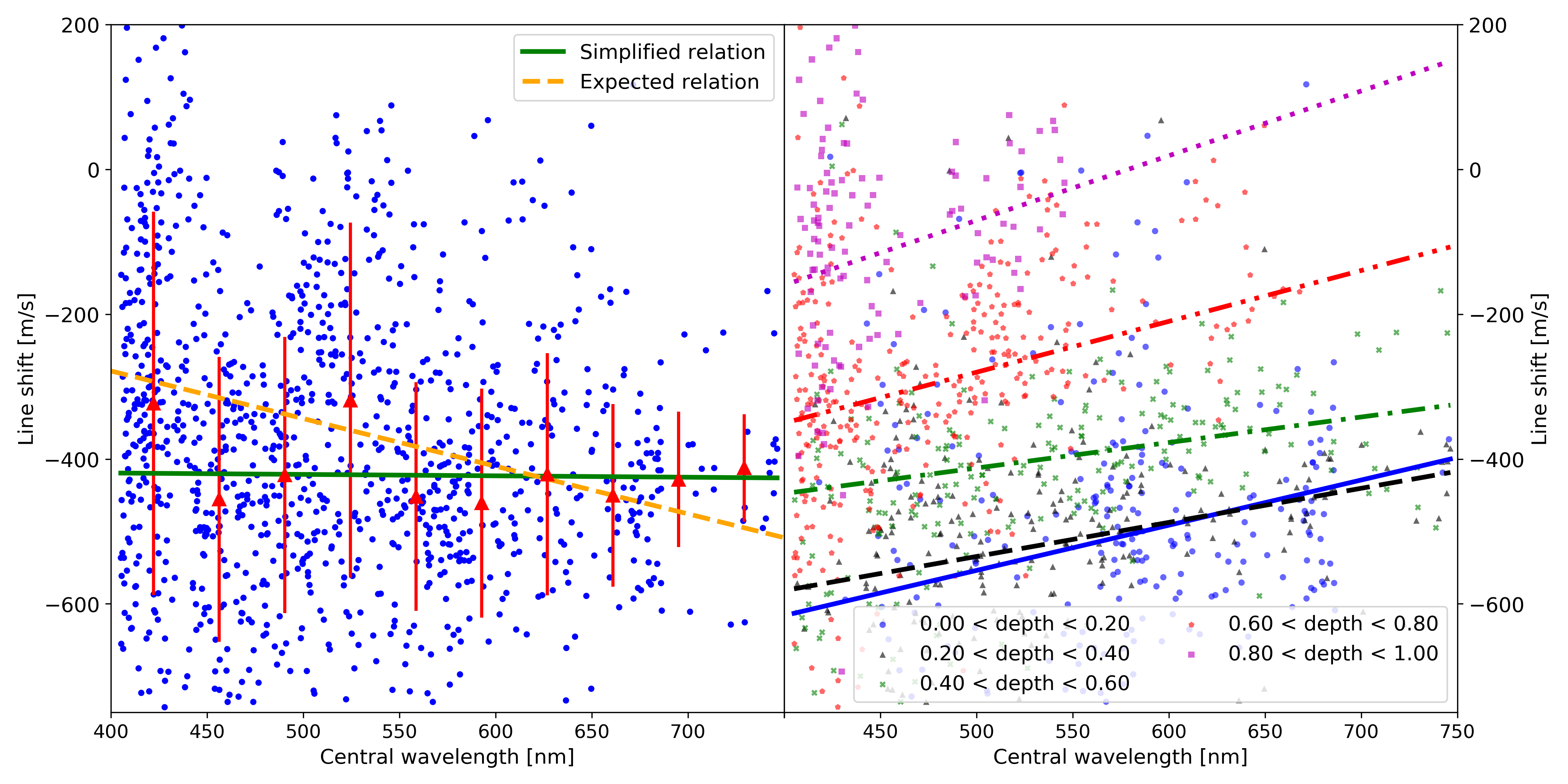}}
\caption{Two approaches to the wavelength dependence of CBS. Left: Simplified attempt at finding the dependence by directly plotting the line shift against the central wavelength (blue dots). Binning the data (red triangles with error bars) and applying a linear fit (green, continuous line) does not reveal any dependence. The orange, dashed line represents the expectation based on Figs. \ref{img:solar_sig} and \ref{img:solar_sig_depth_wav} assuming lines at 400\,nm to have an average depth about 0.7 and at 750\,nm of about 0.2. Right: Depth-binned dependence of line shift on central wavelength. By largely removing the influence of line depth through binning, the actual wavelength effect becomes visible. The slopes of the binned wavelength effect compared to the depth distribution effect (left panel, dashed line) reveal an equal but opposite magnitude, accounting for the missing wavelength dependence in the simplified approach (left panel, solid line).}
\label{img:solar_sig_wav}
\end{figure*}

The scatter remaining in Fig. \ref{img:solar_sig} can be seen in many works. A commonly explored possibility is the central wavelength of the line as the source. For instance, \citetads{1998A&AS..129...41A} have found no such dependence by plotting line shift against central wavelength, as shown in the left panel of Fig. \ref{img:solar_sig_wav} for the data of this work. The problem with this approach is that such a simple plot will generally show no dependence at all. The reason for this is that the line depths are not evenly distributed in wavelength as shown in Fig. \ref{img:solar_sig_depth_wav}, but have a tendency for deeper lines to be more numerous toward shorter wavelengths. With deeper lines showing less blueshift, one would therefore expect less blueshift at shorter wavelengths as is also shown in the left panel of Fig. \ref{img:solar_sig_wav} for the given distribution of lines and solar template signature. To extract this apparently missing correlation from our solar FTS data, the depth dependence was removed by first binning by depth and then by wavelength. The result is plotted in Fig. \ref{img:solar_sig_wav} and shows the required trend to explain the discrepancy. In the following we proceed to quantify how much of the scatter in blueshift for the solar third signature can be attributed to the wavelength effect.\par
\begin{figure}
\resizebox{\hsize}{!}{\includegraphics{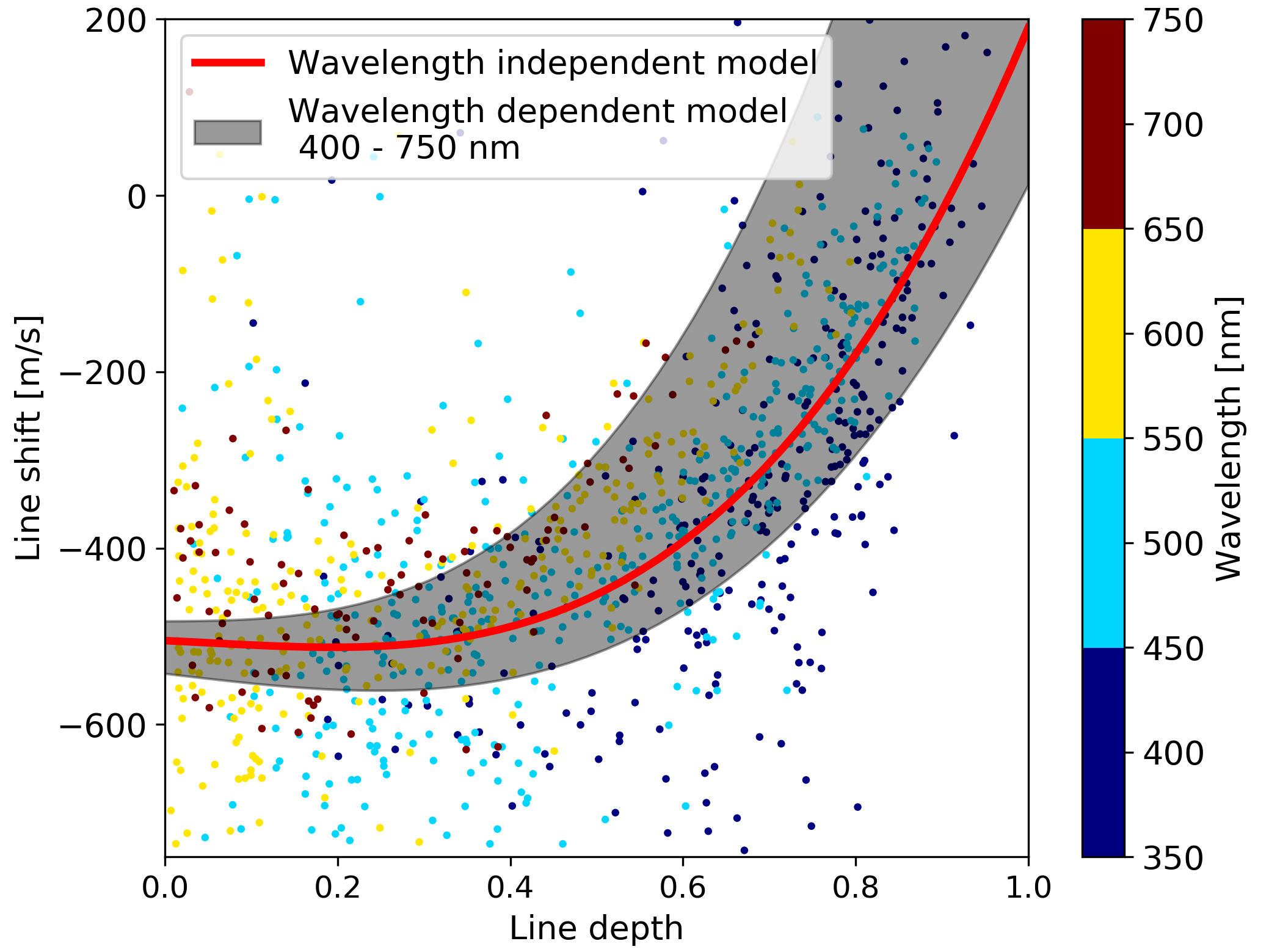}}
\caption{Fitted solar third signature as seen in Fig. \ref{img:solar_sig} (red) that does not take the wavelength dependence into account and the revised model (black shaded) that does. Individual lines are color coded according to their wavelength. The lower boundary corresponds to a wavelength of 400\,nm and the upper to 750\,nm, covering the spectral range of the data. The individual lines' measurements (blue) are shown without their error bars for visibility reasons. The revised model much better captures the spread toward deeper lines.}
\label{img:solar_sig_interval}
\end{figure}

We created a two-dimensional model, linear in wavelength $\lambda$ and of third order in absorption depth $d$, with the third order polynomials in depth defining the coefficients of the linear wavelength dependence.
\begin{align}
v_{\rm conv, \odot}\left(\lambda, d\right) &= m\left(d\right) \cdot \lambda + b\left(d\right),\label{eq:solar_lam_model}\\
m\left(d\right) &= a_{m,1} \cdot d^3 + a_{m,2} \cdot d^2 + a_{m,3} \cdot d + a_{m,4},\label{eq:solar_lam_model_m}\\
b\left(d\right) &= a_{b,1} \cdot d^3 + a_{b,2} \cdot d^2 + a_{b,3} \cdot d + a_{b,4}.\label{eq:solar_lam_model_b}
\end{align}
Before fitting, the data were binned first in depth with 0.1 units width and then in wavelength with ten bins of 35\,nm width spanning 400-750\,nm, similar to Fig. \ref{img:solar_sig_wav}. The model followed Eq. \ref{eq:solar_lam_model} - \ref{eq:solar_lam_model_b} with the resulting fit parameters given in Table \ref{tab:sol_lam_coeff}.
\begin{table}
\centering
\caption{Coefficients for the wavelength dependent, solar third-signature model from Eq. \ref{eq:solar_lam_model} - \ref{eq:solar_lam_model_b}}
\label{tab:sol_lam_coeff}
\begin{tabular}{c c c c}
\hline\hline
Parameter & Value & Parameter & Value\\
\hline
$a_{m,1}$ & 2.04 & $a_{b,1}$ & -123.56\\
$a_{m,2}$ & 0.21 & $a_{b,2}$ & -108.85\\
$a_{m,3}$ & 0.34 & $a_{b,3}$ & -249.90\\
$a_{m,4}$ & 0.17 & $a_{b,4}$ & -610.50\\
\hline
\end{tabular}
\end{table}
The boundaries on the spread expected from the covered wavelength range are shown in Fig. \ref{img:solar_sig_interval}, and seem to encapsulate most of the scatter for the deeper lines. The shallow lines still show a much larger scatter, which is not unexpected for previously explained reasons. The wavelength dependent model also agrees with the independent one, if one inputs the median wavelength of the lines used. In the main part of this work, it is the wavelength independent model from Eq. \ref{eq:template_third_sig} that was used, as explained in Sect. \ref{sec:reference_data}.

\section{Algorithm performance}
\label{sec:apdx_algo_perf}
\subsection{Accuracy of line center}
\label{subsec:acc_line_center}
\begin{figure}
\resizebox{\hsize}{!}{\includegraphics{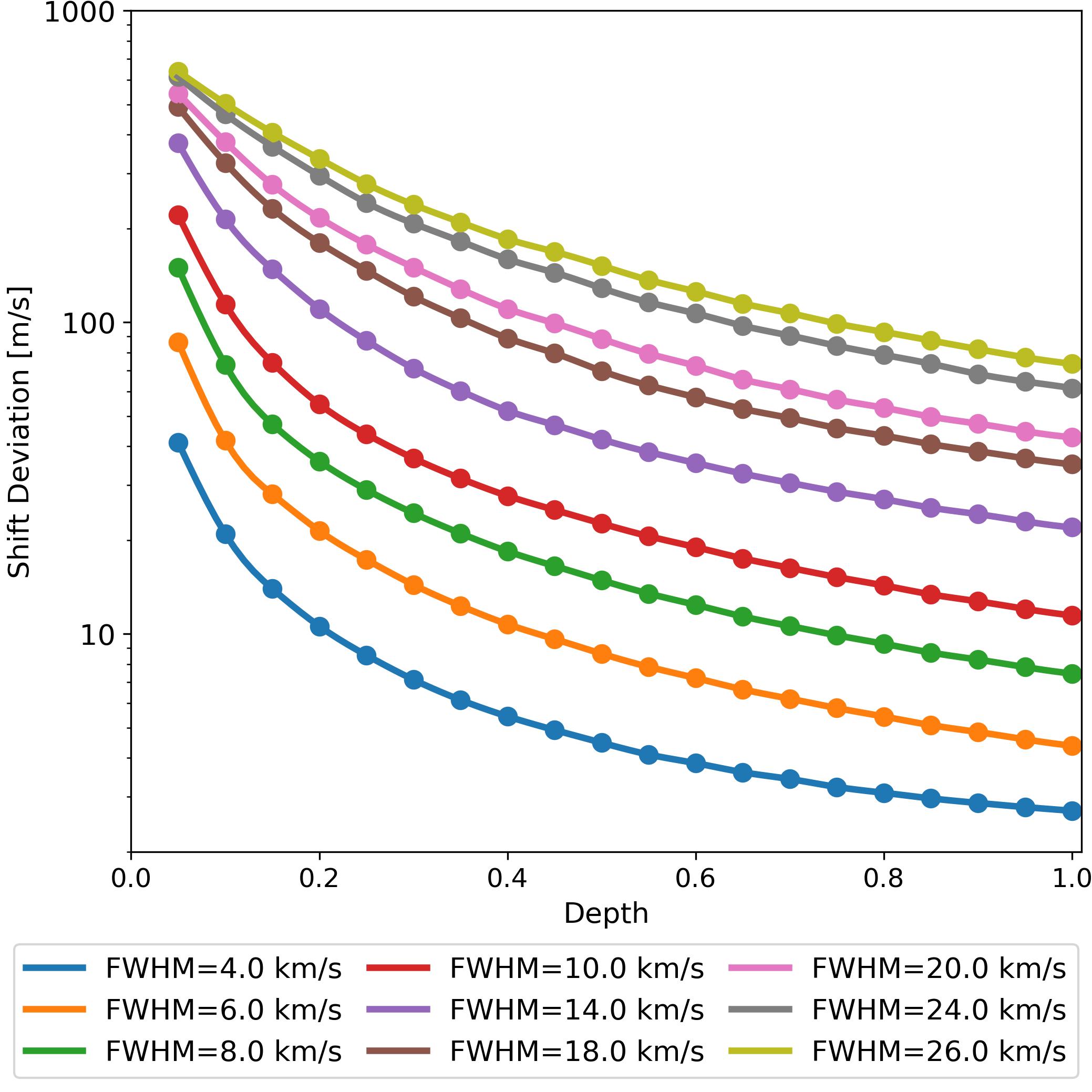}}
\caption{Deviation between the measured center of an artificial, Gaussian absorption line and the true center with Gaussian noise applied and randomized initial measuring points.}
\label{img:acc_line_center}
\end{figure}

Determining the third-signature strength depends on line depth and shift, with the latter, as a first moment, being more difficult to determine exactly. For that reason we tested a set of artificial, Gaussian lines, generated for a resolving power of R = 1,400,000, a S/N of 340, a central wavelength of 600\,nm, and to match a given full-width-at-half-maximum (FWHM) and central depth. The Gaussian noise was applied to 1000 instances of the profile with different random seeds and measurements started at 200 starting points, again randomly distributed with a Gaussian standard deviation of 500\,m\,s$^{-1}$. The starting points were the same on each of the 1000 instances. Figure \ref{img:acc_line_center} shows the resulting absolute deviations between the measured center of the line and the actual center, median averaged over all instances and starting points. The figure shows that for FWHM values that can be expected for the reduced line list, the expected uncertainty in the center of individual lines in the depth range in question is less than 40\,m\,s$^{-1}$, underlining the need for many lines to be available for measurement. It also means that for fast rotating stars with strongly broadened lines, measuring the line centers becomes harder. This can be taken from Fig. \ref{img:acc_line_center} by selecting a fixed depth of a line and moving toward higher equivalent widths, that is, upward to higher deviations.

\subsection{Accuracy of line depth}
\label{subsec:acc_line_depth}
\begin{figure}
\resizebox{\hsize}{!}{\includegraphics{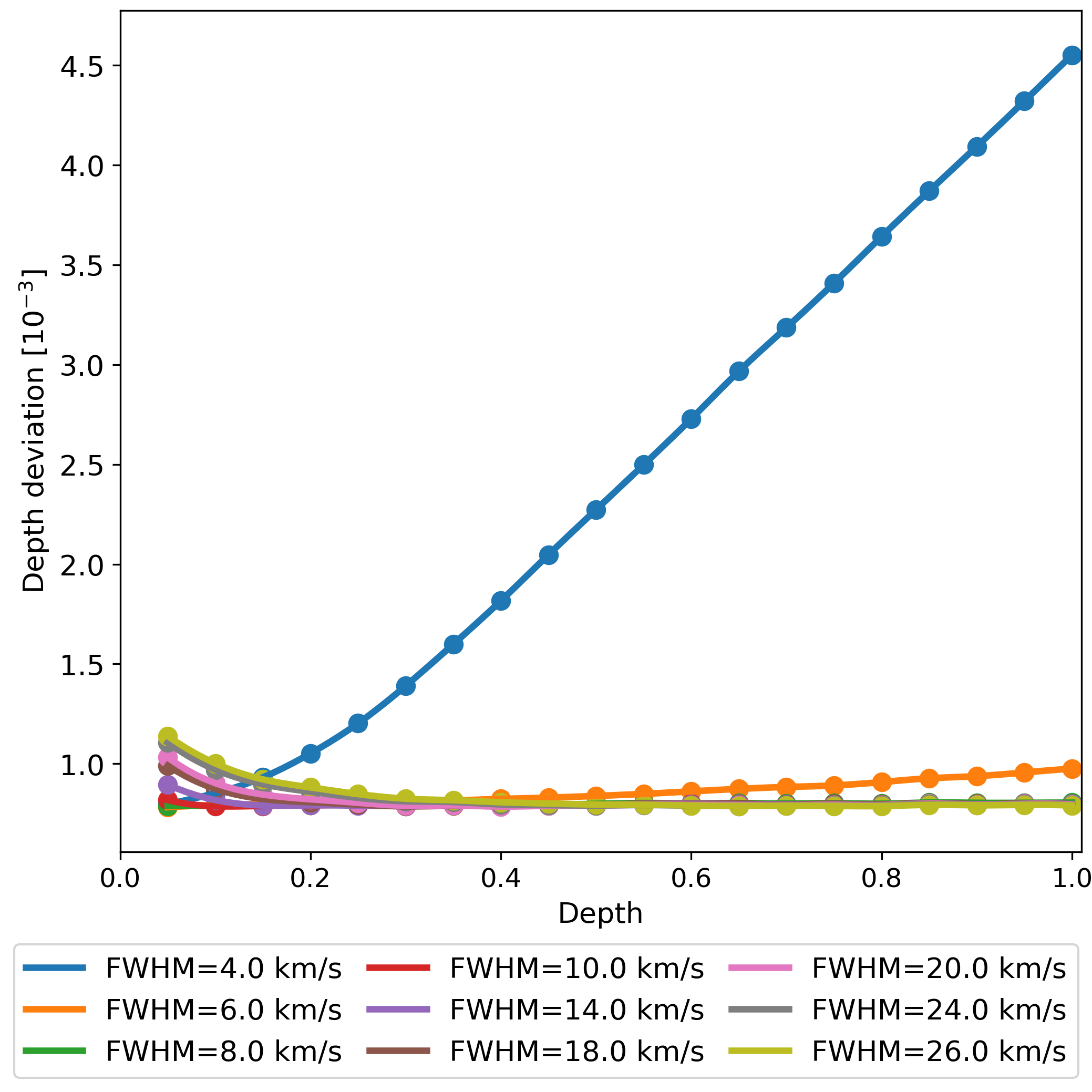}}
\caption{Deviation between the measured depth of an artificial, Gaussian absorption line and the true depth with Gaussian noise applied and randomized initial measuring points.}
\label{img:acc_line_depth}
\end{figure}

Reusing the simulated Gaussians from Sect. \ref{subsec:acc_line_center} also allows for an assessment of the accuracy of the determined line depths. From Fig. \ref{img:acc_line_depth} it can be seen that the deviations are at worst on the order of 4 per mill. Unlike the central shift, the depth not only becomes less accurate for shallow, wide lines but also for very deep ones. This is due to the fixed width of the fitted region that starts to encompass parts of the line where the parabolic approximation breaks down. This could be mitigated by employing a check for an increase in residual toward the wings, indicating a too wide fitting window, or replacing the parabola with a Gaussian function. Both of those would increase the computational overhead while the deviations are small and rare enough to be of no consequence compared to all other sources of error.\par
\begin{figure}
\resizebox{\hsize}{!}{\includegraphics{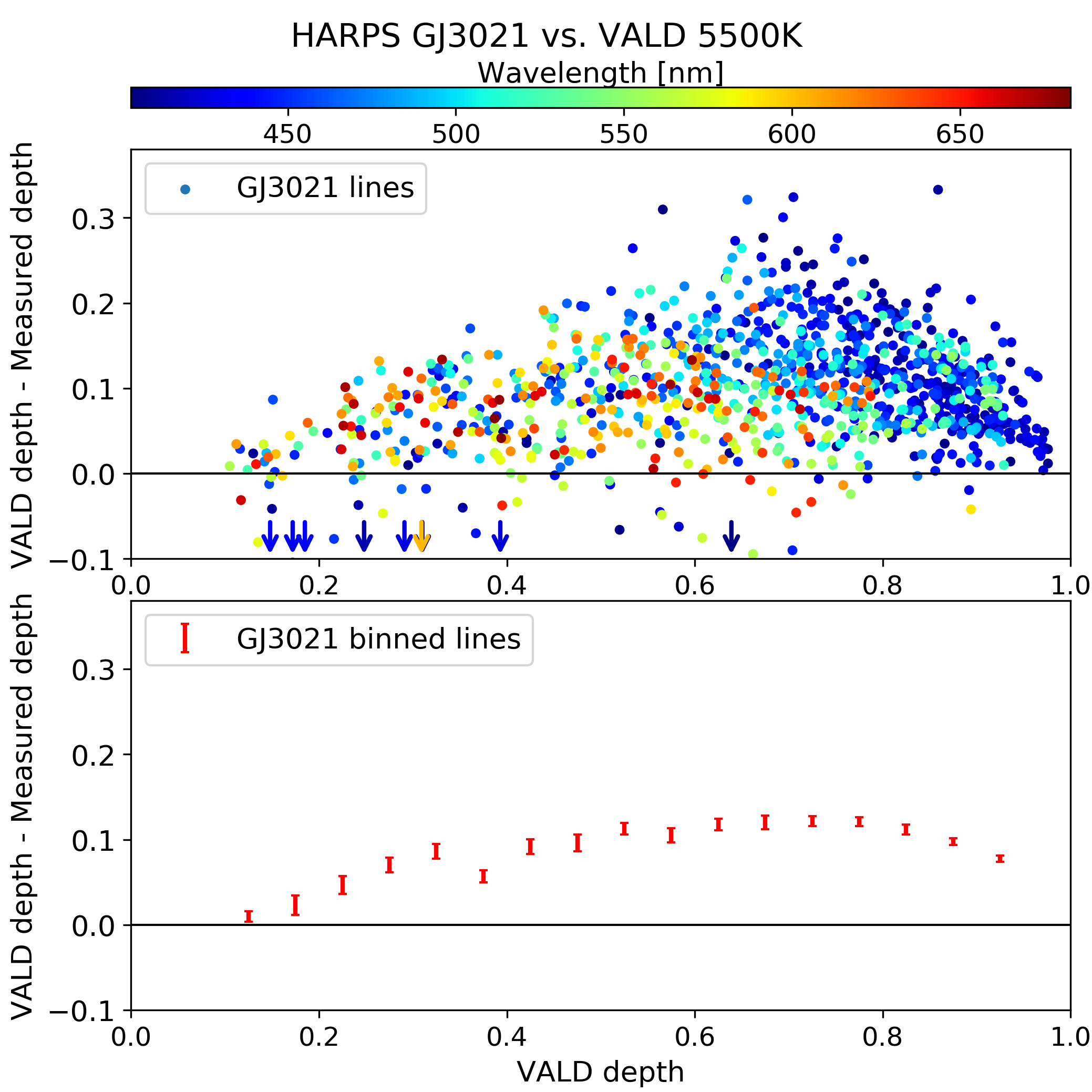}}
\caption{Deviation between VALD line depths and Measured depths from the GJ3021 HARPS spectrum. Top: Deviations of individual lines, color coded for wavelength. Bottom: Binned deviations with uncertainties.}
\label{img:VALD_depth_acc}
\end{figure}

A similar check was performed using the original VALD extractions listed depths and comparing those to measurements of the closest matching star of our HARPS sample. For the 5500\,K case and GJ3021 as the candidate, this is shown in Fig. \ref{img:VALD_depth_acc}. A systematic deviation is clearly visible due to the influence of instrumental and rotational line broadening interacting with the distribution of line equivalent width against absorption depth. Both very shallow and very deep lines show a higher width, relative to their depth, and are less affected by broadening effects, remaining closer to the VALD listed values, while wavelength has no effect. The effect of this systematic is explored in Sect. \ref{subsec:R_effect}, here we focus on the implications of the precision in line depth determination rather than accuracy. Binning the lines in depth following Sects. \ref{sec:reference_data} and \ref{sec:third_sig_fit} reveals binned uncertainties between $3.7\cdot 10^{-3}$ and $11.3\cdot 10^{-3}$ with an average of $7.3\cdot 10^{-3}$. While significantly less precise than the synthetic expectation, this is still well within the acceptable range when compared to the differential RV uncertainties. After conversion to an RV shift (Eq. \ref{eq:template_third_sig}) the depth uncertainty translates to $\lesssim$10\,m\,s$^{-1}$, on par with the expected uncertainty for the RV determination itself. Similar precision can be found when comparing the three remaining line lists from Table \ref{tab:vald_params}.

\subsection{Effects of resolving power}
\label{subsec:R_effect}
\begin{figure}
\resizebox{\hsize}{!}{\includegraphics{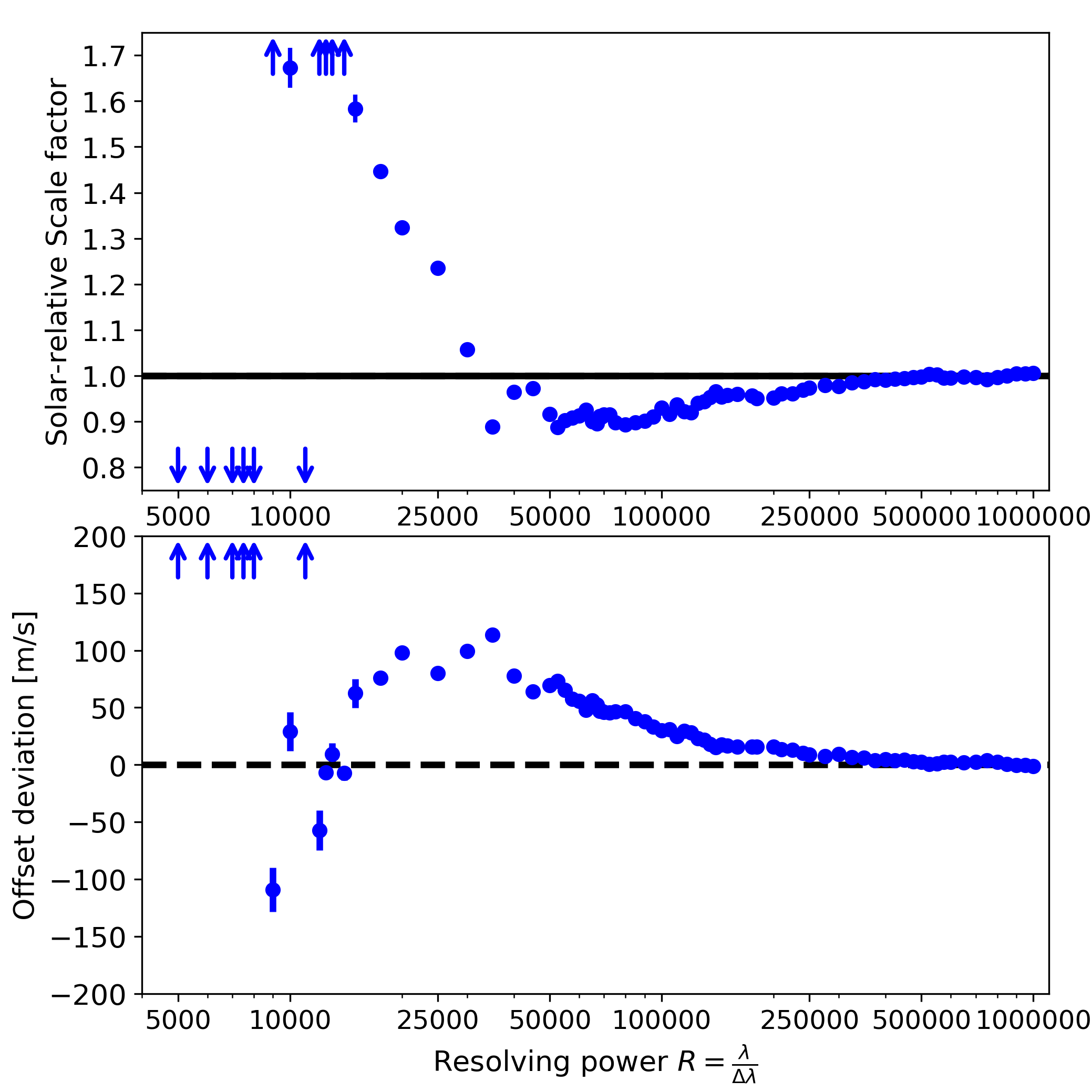}}
\caption{Effect of resolving power, R, on the determined solar scale factor. Top: Scale factor against resolving power, where individual results with different velocity offsets but identical resolving power are median-binned. The horizontal line indicates a scale factor of one, identical to the result from the unconvolved, unshifted FTS. Bottom: Measured velocity offset from the third signature fit, corrected for the synthetic one and binned by R. The dashed line marks zero velocity offset.}
\label{img:solarRTest}
\end{figure}

While HARPS is already an instrument of tremendous resolving power ($R \gtrsim 100,000$), an FTS has $R > 1,000,000$. Effects due to lower resolving power need to be taken into account (see Sect. \ref{sec:reference_data}), especially since Sect. \ref{subsec:acc_line_center} has shown that narrow lines, relative to their depth, are the best targets but also the first to get smeared out in broadened, low-R spectra. To check the effect this has on the determination of blueshift strength, the FTS solar atlas was convolved with a Gaussian broadening kernel to simulate a lower resolving power and Doppler shifted by values spanning $\pm$100\,m\,s$^{-1}$ in 25\,m\,s$^{-1}$ steps. The reduced line list that was used for the HARPS data was applied to the convolved and shifted solar spectrum and the third-signature scale factor determined. This was repeated for a range of resolving powers. The results, median averaged over the offsets, can be seen in Fig. \ref{img:solarRTest}. Resolving powers down to about 50,000 can be corrected for, using correction factors as given in Sect. \ref{sec:reference_data}, as they have close to no effect on the precision of the results, as opposed to the accuracy. The decrease in observed blueshift is due to originally deeper lines, with correspondingly lower blueshift, appearing shallower, thereby lowering the average blueshift in the measured bins. It can be concluded that HARPS' resolving power does not pose a problem in this regard; however, attempting the same with lower-resolution instruments (e.g., $R \lesssim 50000$) leads to a different conclusion. Below a resolving power of $R \approx 10000$ the technique breaks down completely.

\subsection{Effects of stellar rotation}
\label{subsec:vsini_effect}
An analogous analysis to Sect. \ref{subsec:R_effect} was performed for stellar rotation, parameterized by the projected rotational velocity $v\sin i$. We again selected the FTS spectrum degraded to $R \approx 110.000$ and applied rotational broadening from the PyAstronomy package {\tt PyAstronomy.pyasl.asl.rotBroad} for a range of velocities. This reveals a behavior very similar to Fig. \ref{img:solarRTest}. At $R \approx 110.000$, $v\sin i$ leaves the scale factor largely unaffected up to 6\,km\,s$^{-1}$. Similar deviations to $R \approx 25.000$ become visible at velocities of 8\,km\,s$^{-1}$, with intermediate behavior matching as well. In terms of FWHM broadening this is unsurprising, as pure Gaussian broadening at $R \approx 25.000$ and mixed broadening at $R \approx 110.000$ and $v\sin i =$ 8\,km\,s$^{-1}$ both result in FWHM $\approx$ 14\,km\,s$^{-1}$, the apparent upper limit of our method.

\subsection{Effects of signal-to-noise}
\label{subsec:S/N_effect}
\begin{figure}
\resizebox{\hsize}{!}{\includegraphics{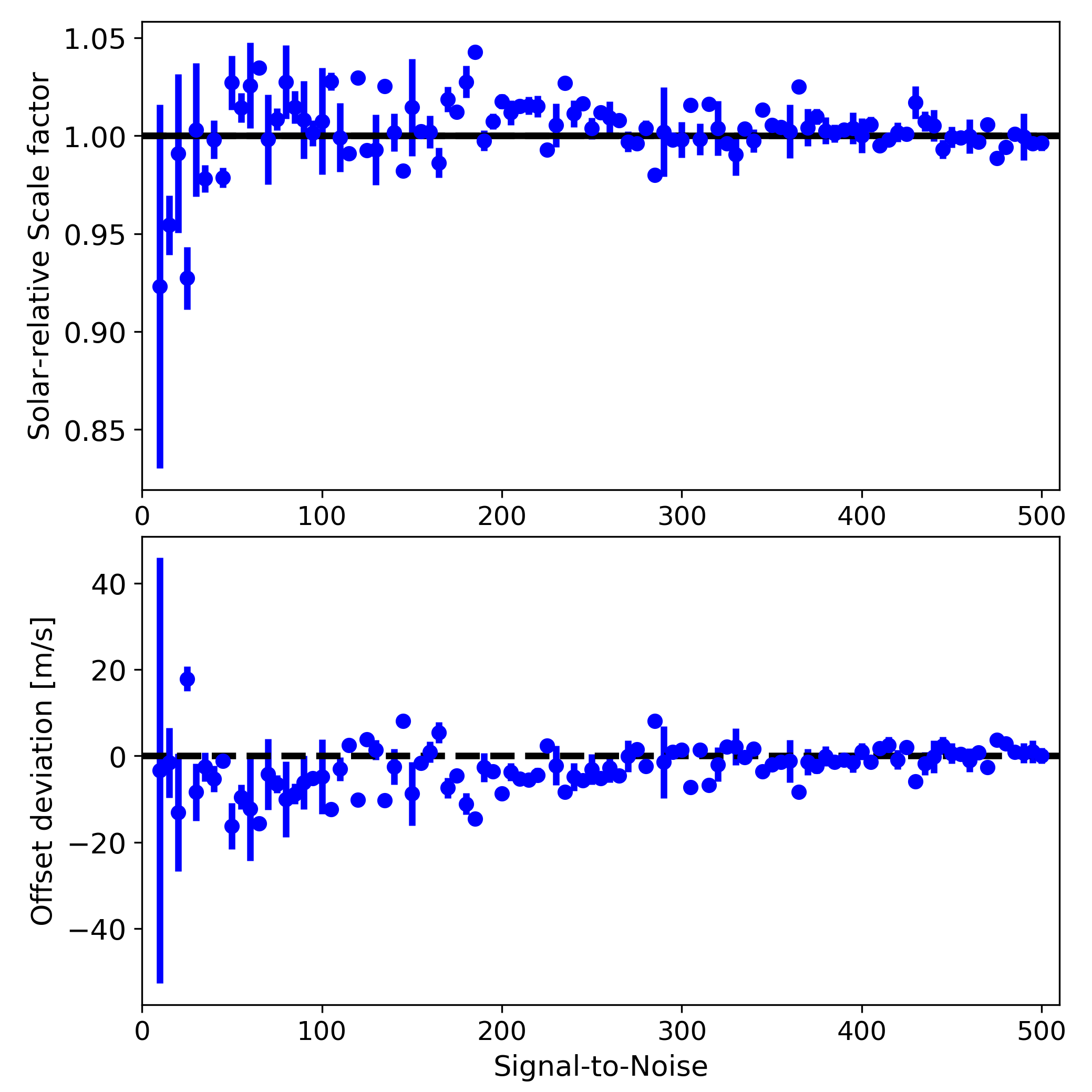}}
\caption{Effect of signal-to-noise ratio on the determined solar scale factor. Top: Scale factor against S/N, individual results with different velocity offsets but identical S/N are median-binned. The horizontal line indicates a scale of one, identical to the result from the undegraded, unshifted FTS. Bottom: Measured velocity offset from the third-signature fit, corrected for the artificial velocity. The dashed line marks zero velocity offset.}
\label{img:solarS/NTest}
\end{figure}

Analogous to the resolving power investigated in Sect. \ref{subsec:R_effect}, the influence of the signal-to-noise ratio can be quantified. Again, the solar FTS was used as the template but instead of being convolved with a broadening kernel, Poissonian noise was added to a specific S/N and the entire spectrum shifted to given offsets, identical to the test for resolving power. The result is shown in Fig. \ref{img:solarS/NTest}. The results in scale and offset are mostly consistent down to S/N$\sim$50. Below that threshold the scale factor drops significantly, which could be corrected for, but also loses precision and the results must be considered unreliable. An S/N above 200 on the other hand seems to be fully unproblematic and gives fully consistent results, with 100 < S/N < 200 remaining acceptable. This reinforces the reliability of the scale factors for all stars in our sample as only a single star shows questionable S/N < 50 and nine stars are 50 < S/N < 100.

\section{Additional figures}
\begin{figure}
\resizebox{\hsize}{!}{\includegraphics{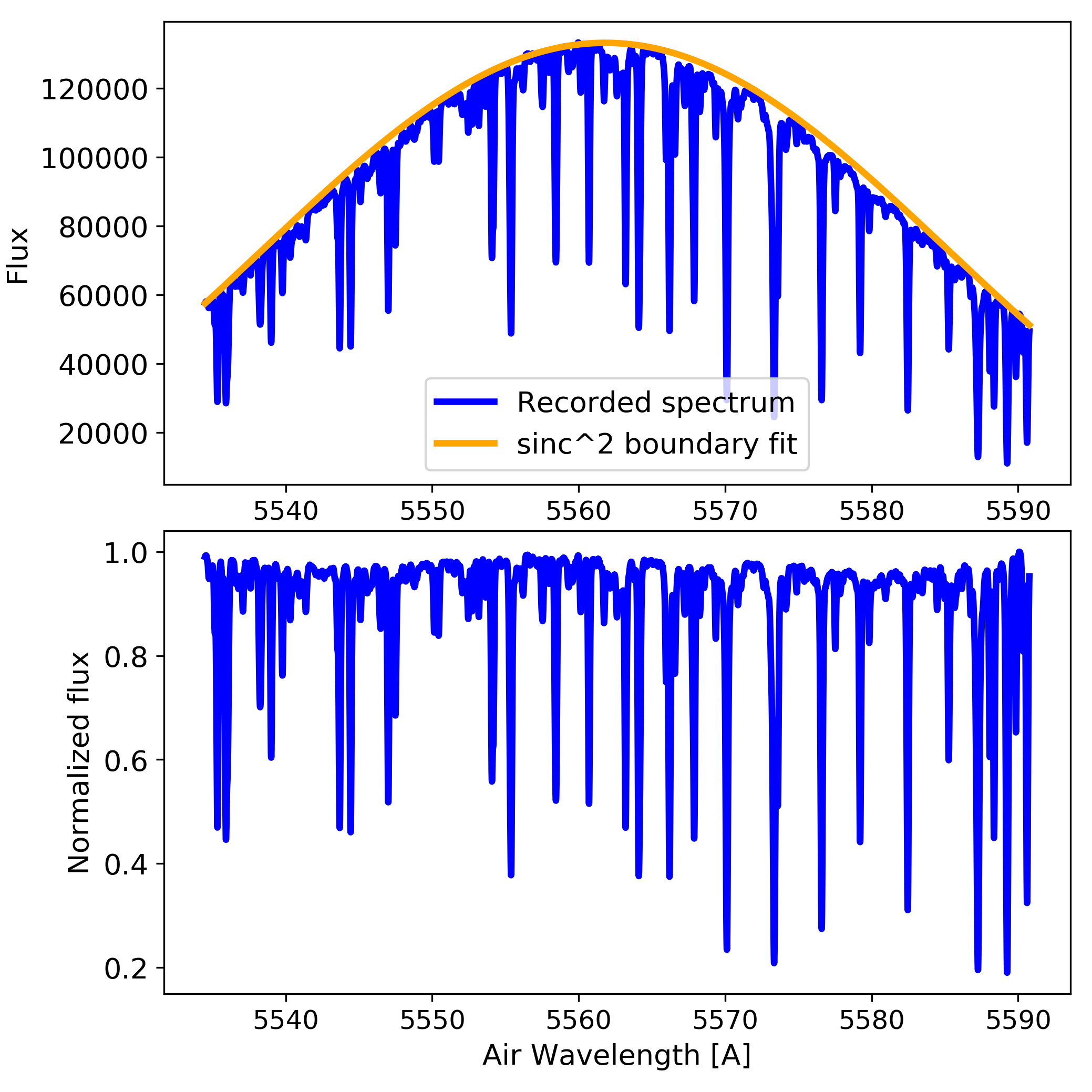}}
\caption{Example for the continuum normalization. Top: Example for a sinc$^2$ boundary fit (orange curve) to the 50th recorded echelle order (blue curve) from GJ4340. Bottom: Normalized flux.}
\label{img:boundfit_example}
\end{figure}

\begin{figure*}
\resizebox{\hsize}{!}{\includegraphics{{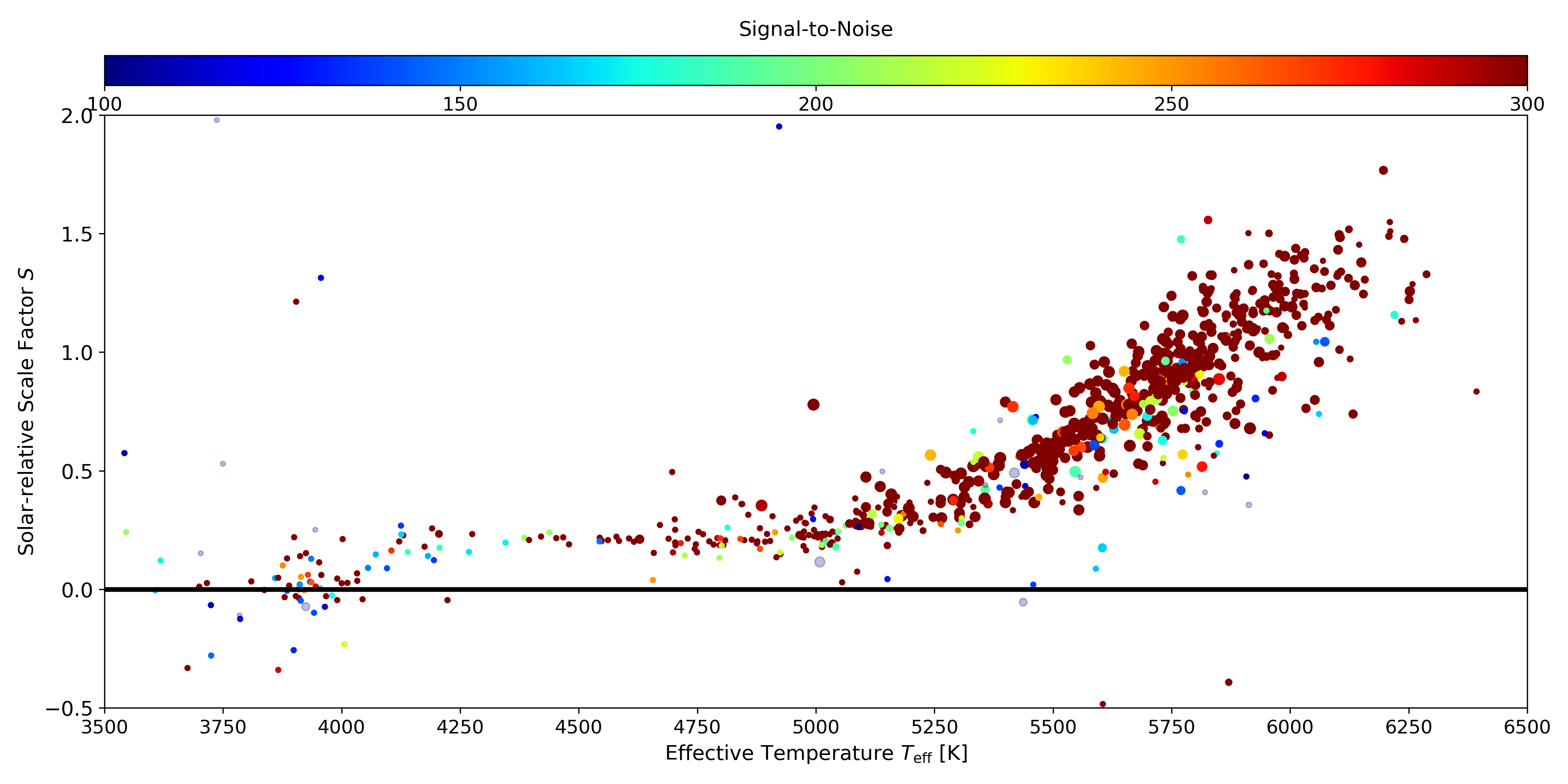}}}
\caption{Third-signature scale for reduced line list with points scaled following the fits $\chi^2$. Smaller points correspond to larger $\chi^2$ values.}
\label{img:apdx_chi2scale_full}
\end{figure*}

In this section of the appendix we provide two figures that expand on points raised in the main text. Figure \ref{img:boundfit_example} shows an example spectral order to illustrate the workings of our sinc$^2$ boundary fit. Figure \ref{img:apdx_chi2scale_full} expands on Fig. \ref{img:scale_Teff} by additionally scaling the marker sizes with their corresponding $\chi^2$ values from the third-signature fit and color coding the S/N of the spectra.
\label{sec:apdx_addFigures}

\section{Tables}
\label{sec:apdx_tables}
For ease of access and a quicker overview, we list in Table \ref{tab:cheatsheet} the expected CBS strengths and velocities for the range of spectral subtypes from F0 to K8 based on our model from Eq. \ref{eq:scale_model}. Table \ref{tab:star_data} lists all stars of our sample with basic parameters important to this work and the determined CBS strengths and uncertainties. The full Table is available via CDS.
\begin{table*}
\centering
\caption{Scale factor cheat sheet following Eq. \ref{eq:scale_model}. Available online at CDS.\label{tab:cheatsheet}}
\begin{tabular}{c c c c}
\hline\hline
Spectral Type & Effective temperature [K] & Scale factor & Velocity [m\,s$^{-1}$]\\
\hline
F0 & 7220 & 6.019 & -2163.1\\
F1 & 7030 & 4.926 & -1770.5\\
F2 & 6810 & 3.844 & -1381.6\\
F3 & 6720 & 3.455 & -1241.6\\
F4 & 6640 & 3.133 & -1125.9\\
F5 & 6510 & 2.657 & -954.8\\
F6 & 6340 & 2.117 & -760.7\\
F7 & 6240 & 1.840 & -661.3\\
F8 & 6170 & 1.664 & -597.9\\
F9 & 6060 & 1.413 & -507.9\\
G0 & 5920 & 1.139 & -409.4\\
G1 & 5880 & 1.069 & -384.3\\
G2 & 5770 & 0.896 & -322.2\\
G3 & 5720 & 0.826 & -297.0\\
G4 & 5680 & 0.774 & -278.2\\
G5 & 5660 & 0.749 & -269.2\\
G6 & 5590 & 0.668 & -240.0\\
G7 & 5530 & 0.605 & -217.5\\
G8 & 5490 & 0.567 & -203.8\\
G9 & 5340 & 0.447 & -160.7\\
K0 & 5280 & 0.409 & -146.9\\
K1 & 5170 & 0.351 & -126.1\\
K2 & 5040 & 0.301 & -108.0\\
K3 & 4830 & 0.254 & -91.1\\
K4 & 4600 & 0.235 & -84.5\\
K5 & 4410 & 0.233 & -83.7\\
K6 & 4230 & 0.232 & -83.3\\
K7 & 4070 & 0.224 & -80.4\\
K8 & 4000 & 0.216 & -77.8\\
\hline
\end{tabular}
\tablefoot{The velocities given assume a line depth of 0.7 (Eq. \ref{eq:template_third_sig}), corresponding to a median solar line, in order to match an expected -350\,m\,s$^{-1}$ at 5800\,K.}
\end{table*}

\longtab[2]{
\begin{landscape}
\begin{longtable}{cccccccccccccc}
\caption{List of all stars with their parameters. Full Version available at CDS.\label{tab:star_data}}\\
\hline\hline
Star & DR2 ID & $T_{\rm eff}$ & Type & $G_{\rm mag}$ & $v\cdot\sin i$ & $P_{\rm rot}$ & $R^\prime_{\rm HK}$ & \#Spectra & S/N & \#Lines & $\chi^2_{\rm P}$ & Scale $S$ & $\sigma_S$\\
\hline
\endfirsthead
\caption{continued.}\\
\hline\hline
Star & DR2 ID & $T_{\rm eff}$ & Type & $G_{\rm mag}$ & $v\cdot\sin i$ & $P_{\rm rot}$ & $R^\prime_{\rm HK}$ & \#Spectra & S/N & \#Lines & $\chi^2_{\rm P}$ & Scale $S$ & $\sigma_S$\\
\hline
\endhead
\hline
\endfoot
GJ551 & 5853498713160606720 & $3296$ & M6 & $8.95$ & $4.70$ & $-$ & $-4.30$ & $387$ & $703$ & $1024$ & $2.54$ & $-0.42$ & $0.17$\\
GJ1224 & 4145870293808914432 & $3302$ & M6 & $11.88$ & $2.20$ & $-$ & $-4.35$ & $6$ & $28$ & $907$ & $446969.04$ & $-0.72$ & $0.25$\\
GJ3379 & 3316364602541746176 & $3317$ & M6 & $9.90$ & $2.67$ & $-$ & $-$ & $16$ & $143$ & $878$ & $199.25$ & $0.59$ & $0.30$\\
GJ285 & 3136952686035250688 & $3542$ & M6 & $9.68$ & $1.26$ & $-$ & $-3.68$ & $7$ & $105$ & $881$ & $33.46$ & $0.39$ & $0.29$\\
GJ3148 & 4971681351020953600 & $3546$ & M5 & $10.78$ & $2.18$ & $-$ & $-$ & $62$ & $206$ & $1038$ & $4.80$ & $0.67$ & $0.12$\\
TYC9379-1149-1 & 5260759175761503232 & $3585$ & M5 & $10.21$ & $3.85$ & $-$ & $-$ & $19$ & $150$ & $1044$ & $12.57$ & $-0.02$ & $0.09$\\
HIP31293 & 5260759175761502208 & $3607$ & M4 & $9.37$ & $1.60$ & $-$ & $-4.86$ & $9$ & $161$ & $1051$ & $120.01$ & $-0.02$ & $0.07$\\
HIP17695 & 3250328209054347264 & $3618$ & M4 & $10.50$ & $1.48$ & $-$ & $-$ & $14$ & $174$ & $883$ & $1280.31$ & $-0.17$ & $0.27$\\
GJ729 & 4075141768785646592 & $3675$ & M5 & $9.13$ & $3.00$ & $-$ & $-4.39$ & $115$ & $556$ & $961$ & $0.36$ & $-0.39$ & $0.22$\\
GJ674 & 5951824121022277632 & $3700$ & M4 & $8.33$ & $0.10$ & $-$ & $-4.77$ & $184$ & $1143$ & $1112$ & $28.92$ & $-0.04$ & $0.05$\\
GJ3643 & 3868080055385587712 & $3703$ & M4 & $11.28$ & $4.00$ & $-$ & $-$ & $21$ & $88$ & $1011$ & $2.05$ & $-0.18$ & $0.11$\\
GJ479 & 6078114541943894016 & $3716$ & M4 & $9.56$ & $0.60$ & $-$ & $-4.61$ & $57$ & $472$ & $1101$ & $915.43$ & $0.01$ & $0.05$\\
GJ3404 & 3127503620545728000 & $3725$ & M4 & $10.89$ & $0.68$ & $-$ & $-$ & $14$ & $109$ & $1048$ & $27.99$ & $-0.08$ & $0.10$\\
GJ552 & 1229089524081628416 & $3725$ & M3 & $9.72$ & $5.60$ & $-$ & $-$ & $8$ & $146$ & $1068$ & $5.94$ & $-0.14$ & $0.06$\\
GJ3508 & 657608666099404544 & $3737$ & M3 & $10.73$ & $8.00$ & $-$ & $-$ & $7$ & $71$ & $972$ & $1860.59$ & $0.37$ & $0.68$\\
HIP116003 & 2391670474561181696 & $3750$ & M4 & $9.92$ & $2.80$ & $-$ & $-$ & $31$ & $-$ & $888$ & $0.21$ & $3.88$ & $1.01$\\
GJ618.4 & 5941478335062267904 & $3785$ & M3 & $10.92$ & $1.16$ & $-$ & $-$ & $8$ & $94$ & $1022$ & $2.12$ & $0.04$ & $0.09$\\
GJ477 & 6131930516518662144 & $3786$ & M3 & $10.16$ & $-$ & $-$ & $-$ & $9$ & $116$ & $1038$ & $0.18$ & $0.16$ & $0.09$\\
GJ382 & 3828238392559860736 & $3810$ & M3 & $8.33$ & $1.72$ & $-$ & $-4.59$ & $32$ & $535$ & $1120$ & $45.27$ & $0.01$ & $0.05$\\
GJ876 & 2603090003484151808 & $3837$ & M5 & $8.88$ & $1.62$ & $-$ & $-5.03$ & $257$ & $1001$ & $1098$ & $15.79$ & $-0.07$ & $0.07$\\
GJ821 & 6885776098199760896 & $3854$ & M3 & $10.02$ & $1.00$ & $-$ & $-$ & $11$ & $155$ & $1048$ & $156.11$ & $-0.04$ & $0.07$\\
GJ218 & 2887328533953284096 & $3860$ & M3 & $9.83$ & $1.66$ & $-$ & $-$ & $9$ & $173$ & $1091$ & $30.17$ & $0.01$ & $0.06$\\
GJ16 & 2752746111688997376 & $3860$ & M3 & $10.00$ & $2.34$ & $-$ & $-$ & $8$ & $156$ & $1054$ & $828.56$ & $0.03$ & $0.07$\\
GJ701 & 4177731838628465664 & $3866$ & M3 & $8.52$ & $2.50$ & $-$ & $-4.98$ & $153$ & $1042$ & $1121$ & $80.72$ & $0.03$ & $0.03$\\
GJ3305 & 3205094369407459328 & $3867$ & M3 & $9.72$ & $3.20$ & $-$ & $-$ & $19$ & $285$ & $952$ & $0.44$ & $0.02$ & $0.14$\\
GJ842 & 6409685433069148160 & $3876$ & M2 & $8.91$ & $-$ & $-$ & $-$ & $10$ & $254$ & $1095$ & $67.06$ & $0.10$ & $0.05$\\
GJ361 & 614543647497149056 & $3880$ & M3 & $9.44$ & $1.43$ & $-$ & $-$ & $101$ & $592$ & $1117$ & $57.23$ & $-0.05$ & $0.04$\\
GJ205 & 3209938366665771008 & $3885$ & M3 & $7.10$ & $2.73$ & $-$ & $-4.63$ & $102$ & $1347$ & $1143$ & $0.95$ & $0.17$ & $0.03$\\
GJ645 & 5971414119713770496 & $3885$ & M3 & $10.56$ & $4.60$ & $-$ & $-$ & $14$ & $141$ & $1034$ & $50.74$ & $0.18$ & $0.07$\\
HIP10395 & 4971496564348094464 & $3889$ & M3 & $9.40$ & $3.10$ & $-$ & $-$ & $26$ & $323$ & $1109$ & $35.83$ & $-0.05$ & $0.05$\\
GJ3778 & 3715722687629983232 & $3899$ & M3 & $11.00$ & $2.71$ & $-$ & $-$ & $20$ & $131$ & $1052$ & $390.00$ & $-0.10$ & $0.08$\\
GJ900 & 2646280705713202688 & $3900$ & K7 & $9.03$ & $5.42$ & $-$ & $-$ & $15$ & $323$ & $1121$ & $2.56$ & $0.26$ & $0.04$\\
GJ9103A & 4741883764414703616 & $3904$ & M3 & $9.79$ & $2.70$ & $-$ & $-$ & $48$ & $383$ & $1102$ & $6.41$ & $-0.01$ & $0.04$\\
GJ855 & 6504705571539381248 & $3904$ & M2 & $9.89$ & $4.10$ & $-$ & $-$ & $35$ & $308$ & $1114$ & $557.29$ & $0.02$ & $0.05$\\
GJ390 & 3767878708888402944 & $3910$ & M3 & $9.27$ & $2.90$ & $-$ & $-$ & $48$ & $483$ & $1129$ & $95.49$ & $-0.02$ & $0.04$\\
GJ1 & 2306965202564506624 & $3910$ & M3 & $7.68$ & $4.80$ & $-$ & $-5.33$ & $48$ & $799$ & $1107$ & $4.13$ & $-0.09$ & $0.04$\\
GJ410 & 3988689609004982784 & $3912$ & M2 & $8.81$ & $1.90$ & $-$ & $-$ & $13$ & $328$ & $1101$ & $159.86$ & $0.08$ & $0.05$\\
GJ510 & 6188331614726319104 & $3912$ & M3 & $10.11$ & $0.53$ & $-$ & $-$ & $13$ & $154$ & $1059$ & $6.53$ & $-0.16$ & $0.06$\\
HD42581 & 2940856402123426304 & $3913$ & M2 & $7.31$ & $2.30$ & $-$ & $-4.64$ & $201$ & $1956$ & $1146$ & $9.59$ & $0.16$ & $0.03$\\
GJ637 & 5805810454276180992 & $3914$ & M3 & $10.44$ & $3.53$ & $-$ & $-$ & $17$ & $142$ & $1046$ & $254.85$ & $-0.05$ & $0.07$\\
GJ173 & 3184351876391975936 & $3914$ & M3 & $9.40$ & $2.35$ & $-$ & $-$ & $16$ & $249$ & $1087$ & $39.31$ & $-0.04$ & $0.05$\\
GJ800 & 6858312737278836736 & $3921$ & M2 & $9.90$ & $0.10$ & $-$ & $-$ & $30$ & $253$ & $1107$ & $213.59$ & $0.04$ & $0.05$\\
GJ563.2 & 6228695270697905152 & $3925$ & M3 & $10.79$ & $2.68$ & $-$ & $-$ & $9$ & $87$ & $1024$ & $2444.54$ & $-0.04$ & $0.09$\\
... & ... & ... & ... & ... & ... & ... & ... & ... & ... & ... & ... & ... & ...\\
\end{longtable}
\end{landscape}
}

\end{appendix}

\end{document}